\documentclass[acmlarge]{acmart}

\acmJournal{IMWUT}
\acmVolume{1}
\acmNumber{1}
\acmArticle{1}
\acmMonth{1}

\setcopyright{none}
\renewcommand\footnotetextcopyrightpermission[1]{}

\acmYear{2026}
\copyrightyear{2026}

\usepackage{caption}
\usepackage{graphicx}
\usepackage{tabularx} 
\usepackage[normalem]{ulem}
\usepackage{makecell} 
\usepackage{xspace}
\usepackage{booktabs}
\usepackage{multirow}
\usepackage{cleveref} 
\usepackage{lipsum}
\usepackage{enumitem}
\usepackage{amsmath}
\usepackage{float}
\usepackage{adjustbox}
\setlist[itemize]{noitemsep, topsep=0pt}
\usepackage{adjustbox}
\usepackage{listings}
\usepackage{subcaption}
\usepackage[normalem]{ulem}
\usepackage{algorithmicx}
\usepackage{algorithm}  
\usepackage{algpseudocode} 
\usepackage{amsmath}       
\usepackage{booktabs}   
\usepackage{pifont}     
\usepackage{xcolor}     
\usepackage{colortbl}   
\usepackage{graphicx}   

\usepackage{amssymb}

\definecolor{Gray}{gray}{0.9} 

\usepackage{soul}
\usepackage{hyperref}
\hypersetup{
    colorlinks=true,
    linkcolor=blue,
    filecolor=magenta,
    urlcolor=cyan,
}

\newcommand{\system}{\textit{VeV}\xspace}

\begin{document}

\title[]{\system: Video-Enhanced Vibration Data Augmentation For Structural Difference}


\author{Joonhee Lee}
\email{neo81389@yonsei.ac.kr}
\affiliation{%
 \institution{Yonsei University}
 \country{Seoul, South Korea}
}

\author{Jiale Zhang}
\email{jiale@umich.edu}
\affiliation{%
 \institution{University of Michigan, Ann Arbor}
 \country{MI, United States}
}

\author{Sungmin Lee}
\email{i.am.sungmin.lee@yonsei.ac.kr}
\affiliation{%
 \institution{Yonsei University}
 \country{Seoul, South Korea}
}

\author{Ruiqing Wang}
\email{ruiqing@illinois.edu}
\affiliation{%
 \institution{University of Illinois Urbana-Champaign}
 \country{IL, United States}
}

\author{Jeffrey Shulkin}
\email{shulkinj@umich.edu}
\affiliation{%
 \institution{University of Michigan, Ann Arbor}
 \country{MI, United States}
}

\author{Yiwen Dong}
\email{yiwen@illinois.edu}
\affiliation{%
 \institution{University of Illinois Urbana-Champaign}
 \country{IL, United States}
}

\author{Pei Zhang}
\email{peizhang@umich.edu}
\affiliation{%
 \institution{University of Michigan, Ann Arbor}
 \country{MI, United States}
}

\author{JeongGil Ko}
\email{jeonggil.ko@yonsei.ac.kr}
\affiliation{%
 \institution{Yonsei University}
 \country{Seoul, South Korea}
}

\begin{abstract}
    Surface-mounted vibration sensing provides non-invasive structural monitoring with high temporal fidelity across diverse ubiquitous computing scenarios such as smart farms, yet structure-borne wave models remain fragile to physical relocation and structural shifts. However, addressing this spatial data discontinuity remains limited by costly manual recalibration, insufficient physical grounding in data-driven synthesis, and substantial sim-to-real gaps in numerical digital twins. To resolve this spatial bottleneck without labor-intensive data recollection, we propose \system{}, a physics-consistent spatial augmentation framework that synthesizes vibration responses at unmeasured target coordinates from sparse reference observations by coupling a monocular Vision-Derived Propagation Possibility Map for structural wave-guiding paths with a Visual Feature Map for transient source kinetics and implicit material dynamics. Evaluated on a 1,237 sensor-hour sow pig and piglet farm across disparate structures, \system{} reduces waveform reconstruction MSE by over 57.5\% compared to data-driven baselines and achieves up to 81.69\% zero-shot classification accuracy, while reaching peak performance within just 30–50 calibration chunks to effectively eliminate cold-start retraining overhead and establish a scalable foundation extending to human-centric floor vibration sensing and ambient intelligence.
\end{abstract}

\settopmatter{printacmref=false} 
\setcopyright{none}
\renewcommand\footnotetextcopyrightpermission[1]{}
\pagestyle{plain}

\setcopyright{none}
\makeatletter
\renewcommand\@formatdoi[1]{\ignorespaces}
\makeatother

\maketitle

\section{Introduction}
\label{sec:intro}

Vibration sensing enables continuous monitoring of physical activities and interactions through vibrations transmitted across structures. Surface-mounted vibration sensors can capture subtle mechanical responses that are difficult to observe through other sensing modalities (e.g., optical cameras due to occlusions, poor lighting, imperceptible micro-deformations, ambient microphones prone to acoustic noise, or coarse motion trackers). These capabilities have enabled diverse ubiquitous sensing applications, ranging from ambient human sensing (e.g., footstep-based localization and non-intrusive activity tracking) to smart agriculture (e.g., monitoring animal behavior via structural vibrations), as well as industrial applications such as machinery management, fault diagnostics, and robotic assembly line handling~\cite{Mirshekari2018Occupant,Dong2024PigSense,Mehamud2022Machine,Gong2024AiroTouch,Sandykbayeva2022VibroTouch, ViObject}.

Despite these advantages, maintaining vibration sensing systems in real-world deployments can impose a substantial data collection burden on operators and researchers. In practice, routine maintenance and repairs frequently shift sensors from their original positions. Because structure-borne elastic waves are intricately shaped by local boundary conditions, even minor sensor relocation or structural discrepancies distort the received waveforms, severely compromising cross-deployment generalizability~\cite{Vavva2008BoundaryConditions,Lee2007SensorLocation,Banos2014SensorDisplacement,Stisen2015SmartDevices}.


\begin{figure}[t]
\centering
    \includegraphics[width=0.95\linewidth]{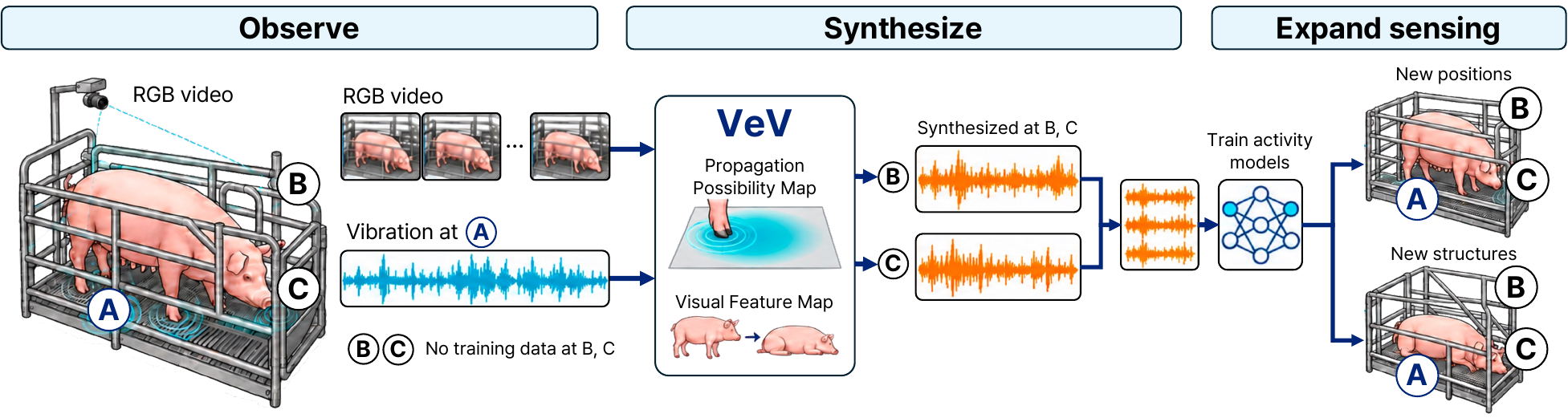}
    \caption{Illustration of \system{}, from Observation to Deployment (Expend sensing). (Part of this illustration was generated using [ChatGPT 6 Astra~\cite{openai2026chatgpt6astra}] to visualize the abstract workflow)}
    \label{fig:teaser}
\end{figure}

This structural sensitivity introduces two distinct deployment challenges: \textbf{intra-structural topology expansion} and \textbf{cross-structural multi-point migration}. First, augmenting an existing deployment with a new sensor (e.g., adding position 4 to positions 1, 2, and 3) creates a lack of historical vibration data at the new sensor location. While existing sensors may have accumulated extensive recordings across diverse activities or operating conditions, these observations cannot directly provide corresponding vibration profiles at the new position, requiring operators to gather and manually annotate new measurements before updating models. Second, when migrating a sensing setup to a structurally similar site where only partial observations are available (e.g., only position 1), the unobserved sensor locations (e.g., positions 2, 3, and 4) lack the baseline and activity profiles necessary for deployment. Repeating data collection and model adaptation at each new site inevitably introduces a prolonged cold-start period, substantially increasing the operational overhead of expanding sensing coverage.

One approach to alleviate such burden is to build foundation models trained on data distributions diverse enough to generalize across deployments. Unfortunately, vibration data are collected in specialized physical environments and cannot be readily obtained from public internet corpora~\cite{Liang2023FewShotVibration,Hassan2024Vibration,Chen2025FederatedTSFM}. Thus, gathering representative data across structural variations, sensor locations, and physical interactions still remains a challenge.

From this perspective, data augmentation offers another way to reduce the amount of new data required for a deployment. Prior work has used data-driven synthesis and physics-based numerical simulations to enrich limited observations~\cite{IwanaUchida2021,YoonJarrettVanDerSchaar2019,YuanQiao2024,MoserJacobsQu1999,WillbergDuczekVivarPerezAhmad2015,WaggWordenBarthorpeGardner2020,SatyanarayanaMelkote2004,YangDengLi2019}. However, learning the distributions of observed signals alone does not determine how vibrations would propagate to an unmeasured sensor location. Furthermore, physics-based numerical simulations, such as the finite element method (FEM), require a detailed digital twin that accounts for material properties, boundary conditions, structural energy dissipation, and sensor contact interfaces which ironically introduces a sim-to-real gap, failing to capture physically grounded vibration responses~\cite{Simoen2015,Ereiz2022,
Cocconcelli2017,Wagg2020}.

Together, these limitations suggest that generating vibration data for a new sensor location requires more than reproducing the characteristics of previously observed signals. Even under the same physical interaction, the received vibration can differ substantially across locations since the wave travels through different structural paths~\cite{vanDer2016,Mitra2016}. Recovering this location-dependent response requires information about the excitation that produces the vibration and the structure through which it propagates. Yet these factors are difficult to characterize in practical deployments, where physical interactions vary over time and the relevant structural properties are only partially known.

To address this challenge, we focus on the complementary information provided by vibration measurements and visual observations. A vibration signal measured at a reference sensor location captures the actual structural response to physical interactions, including effects that are difficult to characterize through numerical simulations. Meanwhile, images reveal structural geometry that can inform possible propagation pathways, while video provides cues about movements and interactions associated with vibration sources. Combining these observations offers an opportunity to estimate vibration responses at unmeasured target locations without requiring a fully calibrated physical model or extensive new vibration measurements.

Motivated by this opportunity, we present \textit{\textbf{\system{}}}, a vision-guided vibration data augmentation framework that synthesizes training data for unmeasured sensor locations by reconstructing their vibration responses from reference measurements and visual observations. Our key idea is to leverage visual observations to characterize structural propagation and vibration-inducing interactions, while using empirical reference vibrations to ground the reconstruction in actual physical responses. Specifically, \system{} introduces a \textit{Vision-Derived Propagation Possibility Map} that captures plausible vibration propagation pathways from visually inferred structural geometry and local continuity, accounting for geometric uncertainty without requiring a calibrated physical model. To further characterize how interactions and material properties influence vibration responses, we develop a \textit{Visual Feature Map for Source Kinetics and Implicit Material Dynamics} that combines inter-frame differential kinetic energies with multi-scale spatial pooling to capture vibration-inducing activities and implicit material characteristics.

Building on these complementary visual representations, \system{} employs a \textit{Spatial-Temporal Wave Propagation Architecture} to reconstruct vibration responses at unmeasured sensor locations. The architecture jointly integrates structural propagation information, source-related dynamics, reference vibration measurements, and the spatial coordinates of reference and target sensors to estimate target vibration responses. By grounding visually inferred propagation and dynamic characteristics in empirical vibration measurements, \system{} enables vibration data augmentation across different sensor placements and structural configurations. The synthesized responses can subsequently augment downstream sensing model training, improving generalization while reducing the need for repeated physical data collection.

We evaluate \textit{\system{}} in a real-world livestock monitoring setting, where animal activities generate structural vibrations across multiple crates. This setting captures key challenges of structural vibration sensing, including unconstrained and dynamic excitations arising from diverse and unpredictable animal behaviors, while providing a relatively controlled environment for systematic evaluation. Our testbed comprises 1,237.023 sensor-hours of vibration data collected from six crates representing two structural types. We evaluate the accuracy of vibration reconstruction at unmeasured sensor locations, its effectiveness in downstream activity recognition, and its generalizability across structural variations. We further investigate the amount of target-location data required for adaptation to assess the practical data collection burden of deploying vibration sensing systems. Experimental results demonstrate that \system{} reduces reconstruction MSE by more than 57.5\% compared to data-driven baselines and achieves up to 81.69\% zero-shot activity recognition accuracy. Furthermore, \system{} reaches peak performance with only 30–50 calibration chunks, substantially reducing the data collection and retraining overhead associated with adapting to new sensor locations.

Specifically, the main contributions of this work are as follows:

\begin{itemize}
    \item \textbf{Vision-Derived Propagation Possibility Map.} We construct a spatial propagation map directly from visual observations to model the geometric structure. This map provides wave-guiding constraints under structural uncertainty such as hidden connections or boundary without requiring precisely calibrated physical models or internal material knowledge.

    \item \textbf{Visual Feature Map for Source Kinetics and Implicit Material Dynamics.} We design a light-weight spatio-temporal representation that pairs multi-scale spatial pooling (GAP, GMP) with inter-frame difference descriptor. This jointly captures transient contact events, motion patterns, and implicit material damping without direct excitation or material measurements.

    \item \textbf{Spatial-Temporal Aware Wave Propagation Architecture.} We design \textit{\system{}}, a novel neural network framework that bridges indirect visual cues and physical vibration dynamics. \system{} fuses visual representations, sensor coordinates, and reference vibrations to accurately reconstruct target-location responses for downstream model training.

    \item \textbf{Zero-Shot Generalization and Rapid Adaptation.} We show that, across extensive real-world evaluations, downstream models trained with \textit{\system{}} achieve superior zero-shot generalization over prior methods, while rapid adaptation with only a small number of calibration chunks (30-50) eliminates cold-start bottlenecks during sensor expansion.
\end{itemize}

\section{Background}
\label{sec:background_motivation}

We review the physical properties of structure-borne vibration, existing approaches to vibration data synthesis, and the role of visual observations in spatial reconstruction.

\subsection{Structure-Borne Vibration and Structural Generalizability}
\label{sec:related_structure}

Structure-borne vibration refers to mechanical vibrations that travel through a solid structure after an external force or physical interaction excites it. Events such as impacts, footsteps, and machinery movements can generate vibrations that propagate through structural elements before reaching a surface-mounted sensor. Consequently, a recorded waveform depends on both the excitation source and the structural environment through which the vibration travels. Even when the same activity occurs, moving a sensor or changing the surrounding structure can alter the measured response.

Several physical mechanisms explain how the structural environment affects a sensor reading. As vibrations travel through a structure, material damping reduces their energy, while boundaries and connections can reflect or transmit waves along different paths. Different propagation paths therefore produce different arrival times, frequency content, and amplitudes at the receiving location. Sensor mounting and contact conditions can introduce additional variation. Together, these factors make the observed signal dependent on where it is measured and how the underlying structure is configured.

In linear elastodynamics, the vibration response can be characterized by how an applied excitation propagates through the structure before reaching a receiving location. Sansalone et al.~\cite{Sansalone1987Transient} demonstrate this relationship using a Green's function formulation to describe impact-induced wave propagation in elastic plates. Their analysis shows how the structural geometry and resulting wave reflections influence the response observed at different locations.

Prior studies in structural health monitoring provide empirical evidence of such spatial dependence. Soman et al.~\cite{soman2019osp} investigate how direct and reflected guided-wave paths affect the coverage provided by different sensor arrangements. Their findings show that boundary reflections can determine whether information from a structural region remains observable. Yang et al.~\cite{yang2017bolted} study Lamb-wave propagation across bolted composite joints and report changes in measured responses under different joint and loading conditions. Both studies illustrate why structural boundaries and connections can modify a received waveform beyond simple measurement noise or uniform amplitude changes.

\begin{figure}[t]
\centering
    \includegraphics[width=0.8\linewidth]{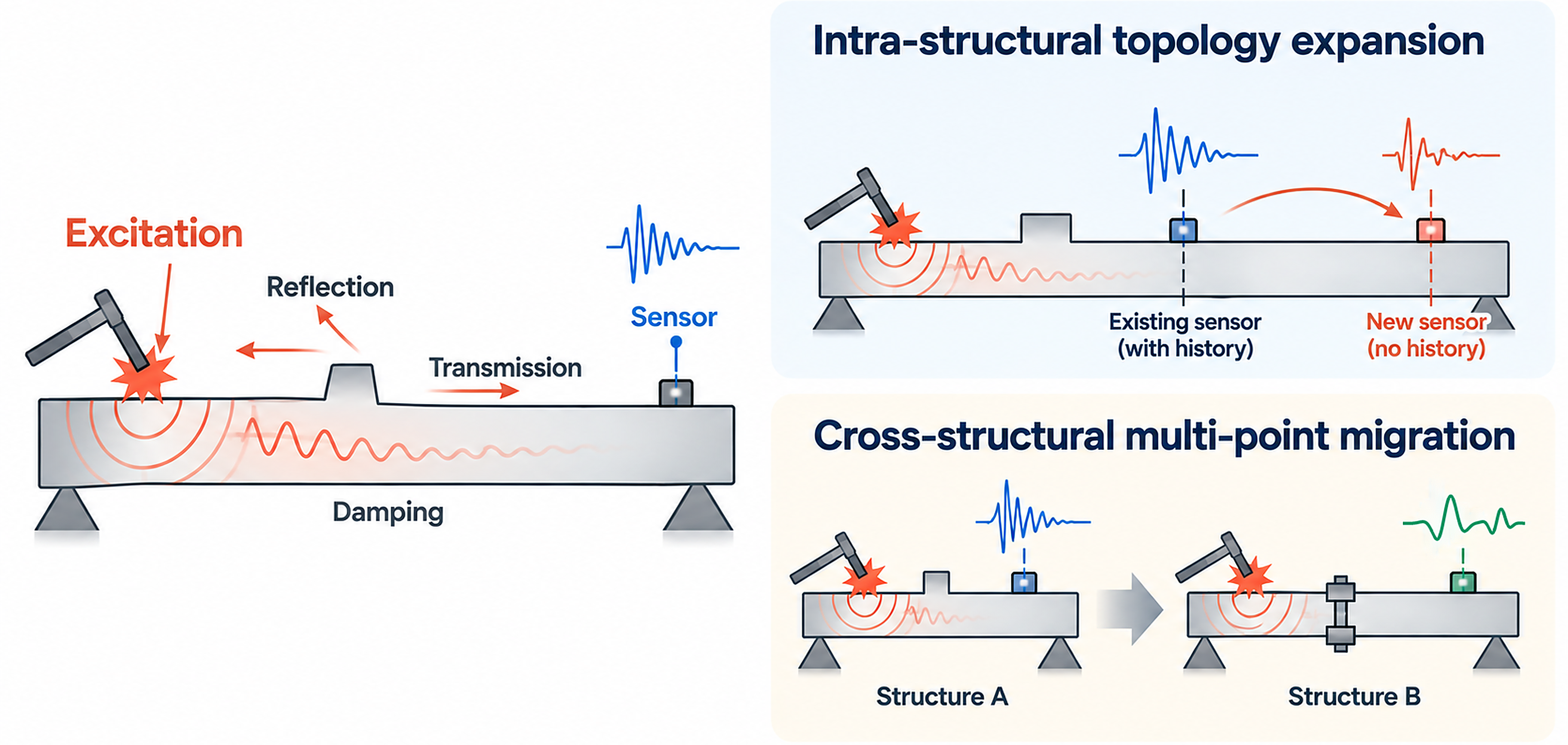}
    \caption{Illustration of intra-structural topology expansion and cross-structural multi-point migration. (This illustration was generated using [ChatGPT 6 Astra~\cite{openai2026chatgpt6astra}] to visualize the abstract workflow)}
    \label{fig:bg_overview}
\end{figure}

Spatial dependence becomes a practical problem when vibration sensing systems are expanded or moved. As illustrated in Figure~\ref{fig:bg_overview}, during \emph{intra-structural topology expansion}, adding a sensor creates a location with no historical measurements, even if other sensors have collected extensive data. During \emph{cross-structural multi-point migration}, apparently similar installations may have different connections and boundary conditions, introducing further changes in the received signal. Worden and Manson~\cite{worden2007ml} likewise emphasize the dependence of data-driven structural monitoring on the conditions represented in the training data. Reusing existing recordings across such changes therefore requires accounting for the response at locations and structures that have not been directly observed.

\subsection{Data Augmentation and Synthesis for Vibration Time Series}
\label{sec:related_augmentation}

Data augmentation helps reduce the need for additional measurements by generating training examples from available recordings or physical models. Existing approaches range from simple signal transformations to learned generative models and numerical simulations. 

\subsubsection{Heuristic Signal Transformations}

Heuristic augmentation produces new training examples by applying predefined transformations to recorded time series. Common operations include jittering, scaling, cropping, temporal warping, and frequency-domain perturbations. Such transformations can improve robustness to changes in amplitude, temporal alignment, or measurement noise. Iwana and Uchida~\cite{iwana2021augmentation} evaluate a broad range of time-series augmentation techniques and show that their benefits depend on the dataset and classification task. Effective augmentation therefore requires a reasonable match between the applied transformation and the variation expected during deployment.

Applying a signal transformation does not directly account for the propagation conditions at a new sensor location. For instance, relocating a sensor may change the arrival times and spectral characteristics of waves reflected from structural boundaries, whereas amplitude scaling or temporal warping modifies an existing waveform without modeling its propagation. Without information about the receiving location and its structural surroundings, such transformations cannot determine which changes would occur along a different path. Their primary benefit therefore lies in improving tolerance to variations represented by the chosen operations, rather than reconstructing a response at an unmeasured physical location. 

\subsubsection{Data-Driven Generative Synthesis}

While heuristic augmentation modifies recorded signals through predefined operations, data-driven generative models learn patterns from the available measurements to synthesize new time series or recover missing observations. TimeGAN~\cite{yoon2019timegan}, for example, combines adversarial and supervised learning to preserve temporal relationships in synthesized sequences. For machinery diagnostics, Ma et al.~\cite{ma2021scgan} investigate sparsity-constrained GAN-based generation of vibration signals. Gu and Ma~\cite{gu2024diffusion} instead use a diffusion-based prior to reconstruct undersampled bearing measurements. Although the tasks differ, these studies demonstrate the value of learned temporal and spectral structure for producing useful vibration-related signals.

The suitability of a generated signal also depends on whether the goal is to diversify training data or to estimate vibration at a particular sensor location. A plausible signal drawn from a learned distribution need not correspond to the response that the same excitation would produce at a specified sensor position. Likewise, recovering missing samples from an existing measurement does not establish how the waveform would change along an unobserved propagation path. Generative synthesis can therefore enrich the available signal distribution, but target-location reconstruction also requires information about the structure and the spatial relationship between measurements.

\subsubsection{Numerical Simulation and Model-Based Virtual Sensing}

Numerical simulation offers a more explicit representation of spatial vibration propagation. Finite-element and related wave simulation methods model the response of a structure using its geometry, material properties, boundary conditions, and applied excitation. Leckey et al.~\cite{leckey2018simulation} compare numerical techniques for guided-wave propagation with experimental measurements, highlighting the importance of faithful physical modeling. Model-based virtual sensing pursues a related objective by estimating responses at unmeasured locations from a limited set of sensors and an assumed structural model. Kullaa~\cite{kullaa2016virtualsensing}, for instance, investigates vibration reconstruction through dynamic substructuring.

Accurate physical modeling, however, requires information that may be unavailable during routine sensing deployments. Hidden connections and contact conditions can be difficult to characterize, while irregular activities introduce forces and excitation locations that are not precisely prescribed. Building or recalibrating a structural model whenever sensors are relocated or installations change may offset the effort saved by synthetic data generation. A practical alternative is to use observations of the actual environment to provide partial structural and interaction-related information, while retaining measured vibration as evidence of the physical response.

\subsection{Visual Observations as Complementary Physical Context}
\label{sec:related_vision}

Visual observations can provide information about structural geometry and physical interactions that is difficult to obtain from sparse vibration measurements alone. Prior studies demonstrate the potential of these cues while also revealing limitations relevant to spatial reconstruction. 

\subsubsection{Visual Measurement of Structural Behavior}

Computer vision has been used to characterize the mechanical behavior of structures without requiring contact measurements at every observed point. Davis et al.~\cite{davis2015visualvibrometry} recover subtle vibration modes from video to infer material-related characteristics. Chen et al.~\cite{chen2015modal} similarly estimate modal properties using high-speed imaging and motion magnification. In structural health monitoring, visual techniques have also been applied to displacement, strain, and vibration-related measurements~\cite{feng2018cvshm}. Such studies show that images contain useful evidence about structural form and observable motion.

Directly measuring mechanical motion through a camera nevertheless differs from predicting the vibration received at another sensor position. Video captures visible surfaces under particular imaging conditions, whereas a contact sensor records the combined response of the structure along its propagation paths. Visual measurements can therefore contribute structural context, but do not on their own determine the waveform produced at an unmeasured receiver.

\subsubsection{Structural Visibility and Geometric Uncertainty}

Inferring propagation conditions from visual geometry also involves uncertainty. An image may show the orientation and continuity of exposed rails or beams, but it cannot reliably identify every hidden joint, support, or material interface. Depth ambiguity and other imaging errors further complicate geometric interpretation, especially outside controlled measurement settings~\cite{zhuang2022cvdeformation}. As a result, visually plausible connections do not necessarily establish how mechanical energy is transmitted through the underlying structure.

Occlusion creates a related distinction between visual visibility and physical observability. Work on non-line-of-sight acoustic sensing has shown that indirect wave measurements can reveal targets outside direct view~\cite{bogerlombard2023nlos}. Although airborne acoustic propagation differs from structure-borne vibration, the broader observation is relevant here because visually hidden regions may still affect a measured physical signal. Spatial reconstruction should therefore use visible geometry as evidence about possible propagation paths while recognizing that unseen structural features can alter the actual response.

\subsubsection{Visual Context for Physical Interactions}

Structural geometry alone cannot explain every difference between vibration recordings. Contact location, movement duration, and impact characteristics influence how an interaction excites the structure, even when its geometry remains unchanged. Video can provide temporal cues about such interactions through visible movement and changes in the scene. Surface appearance can also offer contextual information about the objects involved, although visual features do not directly measure applied forces, stiffness, or damping coefficients.

Combining visual context with physical measurements offers a way to address these complementary limitations. Recorded vibrations reflect the actual excitation and structural response at a known sensing location, while images and video provide partial evidence about propagation pathways, geometric uncertainty, and interaction-related changes. Using both sources of information can support the reconstruction of responses at unmeasured sensor positions without requiring a fully calibrated numerical model.

\section{Preliminary Study}
\label{sec:Prelim}

To characterize the challenges of generalizing vibration-based sensing across sensor locations and structural configurations, we conduct a preliminary study examining the relationship between visual observations, structural vibrations, and downstream sensing performance. Specifically, we investigate how changes in sensor placement and structural configurations affect activity recognition accuracy, and whether these effects can be mitigated by existing data augmentation approaches. Through this analysis, we identify two key challenges in cross-deployment vibration sensing: the limited ability of data-driven models to account for spatially dependent wave propagation and the substantial modeling and calibration requirements of physics-based simulation. These observations motivate the design of \system{}, which combines visual structural information with empirical vibration measurements to support spatial vibration reconstruction.

\subsection{Generalization Problem in Vibration}
\label{subsec:prelim_generalization}

To investigate the sensitivity of vibration-based sensing to sensor placement and structural variations, we compare the generalization performance of vision-only and vibration-only activity recognition models under identical deployment shifts. This comparison allows us to examine whether performance degradation primarily reflects changes in the underlying recognition task or the spatial dependency of structure-borne vibration measurements.

\begin{figure}[t]
    \centering
    \begin{subfigure}[t]{0.31\columnwidth}
        \centering
        \raisebox{-1.9mm}{\includegraphics[width=\linewidth]{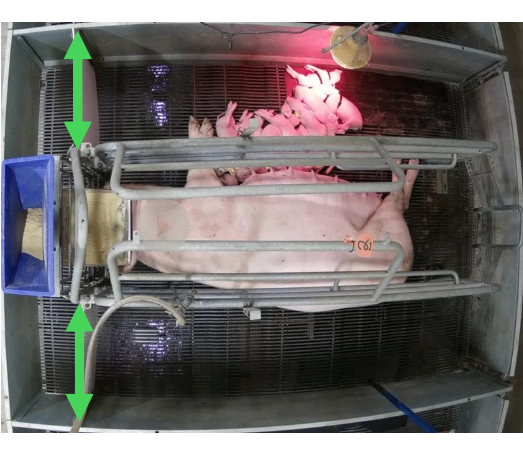}}
        \caption{Seen Crate}
        \label{fig:ref_crate}
    \end{subfigure}\hfill
    \begin{subfigure}[t]{0.333\columnwidth}
        \centering
        \raisebox{-2.8mm}{\includegraphics[width=\linewidth]{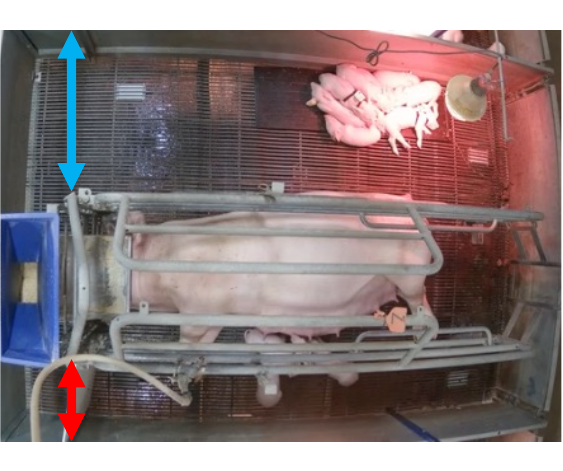}}
        \caption{Unseen Crate}
        \label{fig:cmp_crate}
    \end{subfigure}\hfill
    \begin{subfigure}[t]{0.31\columnwidth}
        \centering
        \includegraphics[width=\linewidth]{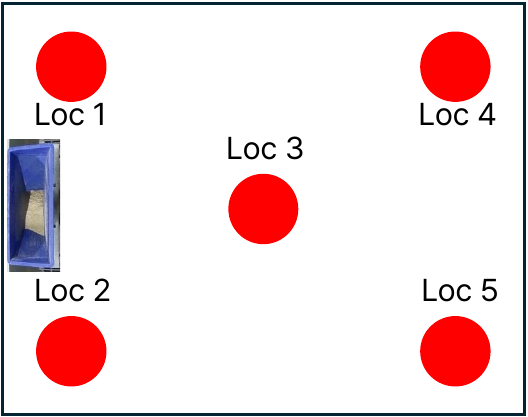}
        \caption{Sensor Location}
        \label{fig:sensor_loc}
    \end{subfigure}
    \caption{Two types of crates with subtle differences, and visualization of sensor locations.}
    \label{fig:crate_n_sensor_visualize}
\end{figure}

We consider sow behavior recognition as a four-class classification task comprising \textit{lying}, \textit{sitting}, \textit{standing}, and \textit{eating+standing}, where the last class represents eating while standing. These classes are selected from the complete farm dataset based on a minimum threshold of 1,000 training instances per class to ensure sufficient representation. As illustrated in Figure~\ref{fig:crate_n_sensor_visualize}, we designate the original deployment (Figure~\ref{fig:ref_crate}) as the \textit{Seen Crate} and a different structural configuration (Figure~\ref{fig:cmp_crate}) as the \textit{Unseen Crate}. Among the five designated sensor positions (Figure~\ref{fig:sensor_loc}), locations 2, 3, and 4 are used during training (\textit{Seen Locations}), while locations 1 and 5 are held out (\textit{Unseen Locations}).

Both vision-only and vibration-only models are trained exclusively on data from the Seen Crate and Seen Locations, and subsequently evaluated under three additional deployment conditions: (i) sensor relocation within the same structure (Seen Crate, Unseen Locations), (ii) structural migration while retaining geometrically corresponding sensor positions (Unseen Crate, Seen Locations), and (iii) combined structural migration and sensor relocation (Unseen Crate, Unseen Locations). To examine whether generalization performance depends on model architecture or capacity, we evaluate four neural architectures (CNN, RNN, LSTM, and Transformer) across three parameter scales of small ($\sim$50k--183k parameters), medium ($\sim$297k--1.02M parameters), and large ($\sim$1.55M--5.24M parameters). While the total parameter counts differ slightly between modalities due to their respective input encoding layers, the backbone architectures are configured with matching parameter budgets.

\begin{table*}[t]
\centering
\caption{Performance Comparison: Vision vs. Vibration under Crate and Sensor Generalization.}
\label{tab:modality_seen_unseen_expanded}
\resizebox{\textwidth}{!}{%
\begin{tabular}{llcccccc}
\toprule
& & \multicolumn{2}{c}{\textbf{Vision Accuracy}} & \multicolumn{4}{c}{\textbf{Vibration Accuracy}} \\
\cmidrule(lr){3-4} \cmidrule(lr){5-8}
& & & & \multicolumn{2}{c}{\textbf{Seen Crate}} & \multicolumn{2}{c}{\textbf{Unseen Crate}} \\
\cmidrule(lr){5-6} \cmidrule(lr){7-8}
\textbf{Arch} & \textbf{Size} & \textbf{Seen Crate} & \textbf{Unseen Crate} & \textbf{Seen Sensor Loc.} & \textbf{Unseen Sensor Loc.} & \textbf{Seen Sensor Loc.} & \textbf{Unseen Sensor Loc.} \\
\midrule
\multirow{3}{*}{CNN} 
& small  & 0.8465 & 0.8381 & 0.8027 & 0.3421 & 0.1802 & 0.1145 \\
& medium & 0.8548 & 0.8471 & 0.7598 & 0.2854 & 0.0957 & 0.0712 \\
& large  & 0.7810 & 0.7911 & 0.7170 & 0.2519 & 0.0801 & 0.0583 \\
\midrule
\multirow{3}{*}{RNN} 
& small  & 0.7323 & 0.7350 & 0.8661 & 0.3812 & 0.1378 & 0.0924 \\
& medium & 0.8465 & 0.8404 & 0.8519 & 0.3340 & 0.0977 & 0.0681 \\
& large  & 0.8199 & 0.8135 & 0.8036 & 0.3625 & 0.2375 & 0.1530 \\
\midrule
\multirow{3}{*}{LSTM} 
& small  & 0.8455 & 0.8410 & 0.8164 & 0.4128 & 0.3267 & 0.2114 \\
& medium & 0.7835 & 0.7901 & 0.8282 & 0.3205 & 0.1015 & 0.0743 \\
& large  & 0.8169 & 0.8347 & 0.9040 & 0.4571 & 0.2488 & 0.1706 \\
\midrule
\multirow{3}{*}{Transformer} 
& small  & 0.8046 & 0.8201 & 0.8317 & 0.3952 & 0.1010 & 0.0768 \\
& medium & 0.8278 & 0.8296 & 0.7539 & 0.3418 & 0.1600 & 0.1129 \\
& large  & 0.8278 & 0.8296 & 0.8071 & 0.4863 & 0.7258 & 0.4135 \\
\bottomrule
\end{tabular}
}
\end{table*}

Table~\ref{tab:modality_seen_unseen_expanded} summarizes the classification performance of both modalities across the evaluated deployment conditions. Vision-based models exhibit relatively stable performance across structural configurations, achieving accuracies of $73.23\%$-$85.48\%$ on the Seen Crate and $73.50\%$-$84.71\%$ on the Unseen Crate. Across all evaluated architectures and parameter scales, the absolute accuracy difference remains below 1.6 percentage points. This suggests that visual representations of animal activities are comparatively insensitive to the structural variations considered in our study. Since cameras capture observable body postures and movements, the relevant visual features remain largely consistent even when the underlying vibration propagation characteristics change.

In contrast, vibration-based models exhibit substantial performance degradation under both spatial and structural shifts. Under the original training configuration, vibration models achieve accuracies of up to $90.40\%$ (LSTM-large) and $86.61\%$ (RNN-small). However, relocating the sensor within the same crate reduces accuracy to $25.19\%$-$48.63\%$, despite the underlying activity classes and structural configuration remaining unchanged. Migrating to the Unseen Crate further challenges generalization, with most models achieving only $8.01\%$-$32.67\%$ accuracy even at geometrically corresponding sensor positions, although Transformer-large retains $72.58\%$ accuracy. When both structural configuration and sensor location change simultaneously, accuracy decreases to $5.83\%$-$21.14\%$ for most models, with Transformer-large achieving $41.35\%$.

These results highlight the strong dependency of structure-borne vibration measurements on both sensor location and structural configuration. Unlike visual observations, which primarily capture the activity itself, vibration signals reflect the combined effects of source excitation and the propagation characteristics between the source and sensor. Conceptually, the observed vibration can be described through a location-dependent transfer operator $g_{s \to r}$, where $s$ and $r$ denote the excitation source and sensor locations, respectively. Relocating the sensor from $r$ to $r'$ changes the corresponding transfer operator to $g_{s \to r'}$, while structural variations further alter the underlying propagation characteristics. Consequently, identical activities can produce substantially different vibration measurements across deployment configurations.

Importantly, increasing model capacity does not consistently overcome these variations. Despite certain larger architectures exhibiting improved robustness, substantial performance degradation persists across the evaluated model families and parameter scales. This suggests that the generalization challenge cannot be addressed solely through model scaling, but requires explicit consideration of the spatial and structural factors influencing vibration propagation. These findings motivate \system{}, which leverages visually inferred structural information and reference vibration measurements to reconstruct vibration responses at unmeasured sensor locations, enabling downstream models to adapt to spatial and structural variations without extensive additional data collection.

\subsection{Failure Analysis of Conventional Augmentation Paradigms}
\label{sec:prelim_augmentation_failure}

Given the substantial generalization degradation observed in vibration-based sensing, data augmentation presents a potential approach to adapting sensing models to new deployment environments. We investigate whether existing data generation techniques can synthesize vibration responses at unmeasured sensor locations and improve downstream activity recognition. Specifically, we consider a cross-deployment scenario in which a classifier initially trained on the \textit{Seen Crate, Position 1} is adapted using synthetic vibration responses generated for the \textit{Unseen Crate, Position 2}. Consistent with Subsection~\ref{subsec:prelim_generalization}, we evaluate performance on the four-class sow behavior recognition task.

We examine four representative augmentation paradigms, encompassing direct data-driven synthesis, transfer learning from pretrained audio models, video-conditioned generation, and physics-based numerical simulation. For learning-based approaches that use reference vibration measurements, we provide the raw vibration waveform together with the spatial coordinates of the reference and target sensor locations, encoded through coordinate embedding layers. These models predict the short-time Fourier transform (STFT) representation of the target vibration response. To facilitate adaptation of pretrained architectures, we employ a two-stage training procedure that first optimizes the newly introduced input and output projection layers while freezing the backbone, followed by end-to-end fine-tuning. For comparison, we also report a \textbf{Real-Data Upper Bound}, obtained by training the downstream classifier directly on actual vibration measurements from the corresponding target environment. This provides a reference for assessing the performance gap between synthetic and real target-domain training data.



\begin{figure}[t]
    \centering
    \begin{subfigure}[t]{0.48\columnwidth}
        \centering
        \includegraphics[width=\linewidth]{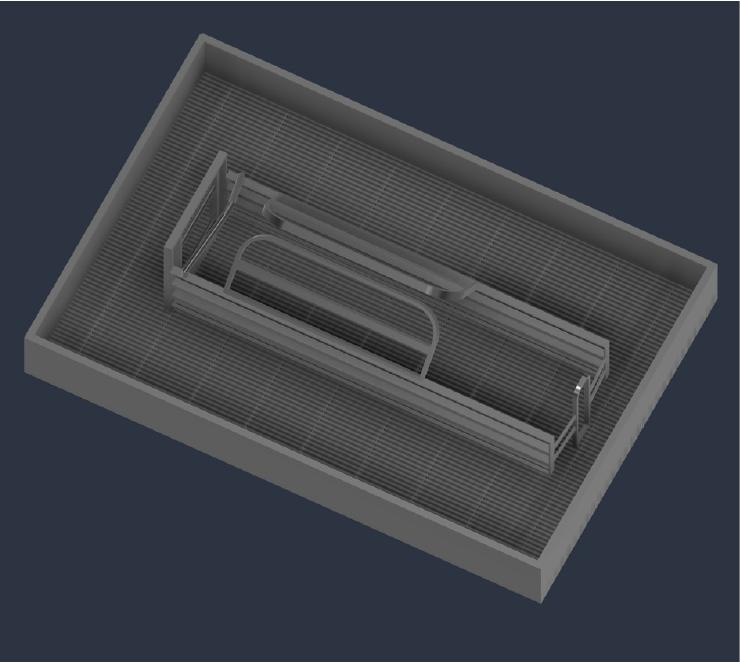}
        \caption{Model of Seen Crate}
        \label{fig:ref_crate}
    \end{subfigure}\hfill
    \begin{subfigure}[t]{0.48\columnwidth}
        \centering
        \includegraphics[width=\linewidth]{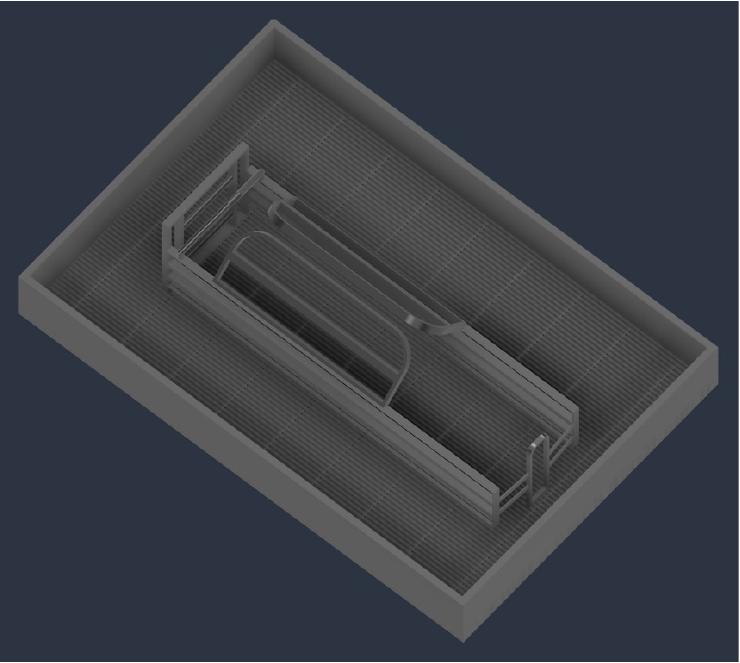}
        \caption{Model of Unseen Crate}
        \label{fig:sensor_loc}
    \end{subfigure}
    \caption{3D CAD and FEM simulation setup constructed based on the visual layout of the Unseen Crate in Figure~\ref{fig:cmp_crate}.}
    \label{fig:fem_modeling}
\end{figure}

\begin{enumerate}
    \item \textbf{Naive Data-Driven Synthesis:}
    We employ a sequence-to-sequence LSTM architecture to learn the mapping between reference and target vibration responses. The model receives the raw vibration waveform from the reference sensor together with the three-dimensional coordinates of the reference and target locations. An input projection layer encodes these features, while an output projection layer predicts the target STFT representation. Following the two-stage training procedure, the model learns spatial vibration mappings from available training pairs on the Seen Crate and subsequently synthesizes responses for the target sensor position on the Unseen Crate.

    \item \textbf{Pretrained Audio Encoder Transfer:}
    Since acoustic and structural vibration signals share oscillatory characteristics, we investigate whether representations learned from large-scale audio pretraining can facilitate spatial vibration reconstruction. We adopt two representative pretrained audio encoders, \textbf{WavLM}~\cite{Chen_2022} and \textbf{BEATs}~\cite{chen2022beats}, and extend them with spatial coordinate embeddings and an output projection head for target STFT prediction. The coordinate embeddings provide information about the reference and target sensor locations, while the pretrained backbone extracts temporal signal representations. We first train the newly introduced projection layers with the backbone frozen, followed by end-to-end fine-tuning using available vibration pairs from the Seen Crate. The adapted models are then used to synthesize vibration responses at the target location on the Unseen Crate.

    \item \textbf{Video-to-Audio Multimodal Generation:}
    To examine whether visual observations alone provide sufficient information for vibration synthesis, we adapt \textbf{MMAudio}~\cite{cheng2025mmaudio}, a video-conditioned audio generation framework. Unlike the preceding approaches, this model does not receive reference vibration measurements. Instead, it generates target vibration representations from visual sequences and the queried sensor coordinates. We introduce a spatial coordinate projection layer to incorporate the target sensor location into the conditioning stream and an output projection head to predict target STFT representations. Following the same two-stage adaptation strategy, we first optimize the newly introduced layers before fine-tuning the multimodal backbone.

    \item \textbf{Physics-Based Finite Element Method (FEM) Simulation:}
    We investigate numerical wave propagation using a finite element model implemented with SfePy~\cite{Cimrman_2019}. As illustrated in Figure~\ref{fig:fem_modeling}, we construct a three-dimensional CAD approximation of the Unseen Crate and identify potential excitation sources by sampling 100 candidate impact locations across the crate floor. For each candidate, we simulate the resulting structural vibration response at Position 1 and compare it against the observed reference waveform. The candidate producing the closest match is selected as the estimated excitation location. Using this estimated source, we subsequently simulate wave propagation toward Position 2 to generate synthetic target vibration responses.
\end{enumerate}

For each augmentation paradigm, we fine-tune the downstream classifier using synthetic vibration responses at Position 2 and evaluate its performance on real target-domain measurements. Table~\ref{tab:augmentation_failure} summarizes the results. The ``Naive Data-Driven'' approach exhibits substantial degradation under deployment shifts, achieving only $21.94\%$ accuracy in the Unseen Crate and Unseen Location setting. This suggests that waveform transformations learned from source-domain measurements fail to generalize to unseen structural propagation characteristics. The ``Audio Encoders'' (WavLM and BEATs) demonstrate comparatively stronger robustness, occasionally achieving higher accuracy on unseen crates than on seen configurations. Nevertheless, their performance remains substantially below the Real-Data Upper Bound, indicating that general-purpose audio representations alone cannot adequately capture location-dependent structural vibrations.

The ``Video-to-Audio'' approach achieves only $4.96\%$ accuracy, highlighting the limitations of reconstructing structural vibrations solely from visual observations. Although video captures visible movements and potential excitation events, it lacks direct measurements of the resulting structure-borne waveforms, which depend on contact conditions, material properties, and propagation pathways. Similarly, ``Physics Simulation'' yields $0.00\%$ downstream accuracy, demonstrating the difficulty of generating effective augmentation data using numerical models under uncertain excitation conditions. In particular, irregular animal-induced impacts and incomplete knowledge of structural parameters introduce discrepancies between simulated and measured responses, limiting the effectiveness of simulation-based augmentation.

Collectively, these results suggest that neither data-driven generation nor numerical simulation alone provides sufficient generalization across structural configurations. Instead, effective vibration augmentation requires combining empirical vibration measurements with visually inferred structural and source-related information to guide spatial reconstruction.

\begin{table}[t]
\centering
\caption{Downstream classification accuracy across varying combinations of seen/unseen crates and sensor locations.}
\label{tab:augmentation_failure}
\begin{tabular}{lcccc}
\toprule
\multirow{2.5}{*}{\makecell[l]{\textbf{Augmentation}\\\textbf{Paradigm}}} & \multicolumn{2}{c}{\textbf{Seen Crate}} & \multicolumn{2}{c}{\textbf{Unseen Crate}} \\
\cmidrule(lr){2-3} \cmidrule(lr){4-5}
& \textbf{Seen Loc} & \textbf{Unseen Loc} & \textbf{Seen Loc} & \textbf{Unseen Loc} \\
\midrule
\makecell[l]{Classifier Upper Bound\\{\small(Classifier w/ real data)}} & 90.40\% & 91.42\% & 90.96\% & 89.87\% \\
\addlinespace[0.4em]
\makecell[l]{(1) Naive Data-Driven\\{\small(Classifier w/ LSTM based synthetic data)}} & 38.05\% & 28.35\% & 42.20\% & 21.94\% \\
\addlinespace[0.4em]
\makecell[l]{(2-a) Audio Encoder \\{\small(Classifier w/ WavLM based synthetic data)}} & 52.74\% & 53.34\% & 63.03\% & 54.33\% \\
\addlinespace[0.2em]
\makecell[l]{(2-b) Audio Encoder \\{\small(Classifier w/ BEATs based synthetic data)}} & 59.23\% & 61.73\% & 69.91\% & 63.07\% \\
\addlinespace[0.4em]
\makecell[l]{(3-a) Video-to-Audio \\{\small(Classifier w/ MMAudio based synthetic data)}} & 5.12\% & 4.97\% & 4.12\% & 4.96\% \\
\addlinespace[0.4em]
\makecell[l]{(4) Physics Simulation \\{\small(Classifier w/ FEM method based synthetic data)}} & 0\% & 0\% & 0\% & 0\% \\
\bottomrule
\end{tabular}
\end{table}

These findings motivate three key design principles for \system{}. First, vibration reconstruction must be grounded in empirical 
reference measurements, as visual observations alone cannot reliably capture the resulting structural responses. Second, visually inferred structural geometry and propagation pathways should complement vibration measurements to account for spatial and structural variations that data-driven models struggle to generalize across. Third, rather than relying on precisely calibrated physical models, reconstruction should incorporate source-related dynamics and structural uncertainty through soft spatial priors. Together, these principles guide the design of \system{}, enabling spatial vibration reconstruction through the integration of empirical measurements and visual context.


\section{Design of \system{}}
\label{sec:System}

Motivated by the findings of our preliminary study, we develop \system{} to reconstruct vibration responses at unmeasured sensor locations or environments from reference measurements and visual observations. This section first formulates the spatial reconstruction problem and then presents the physical considerations underlying the system design.

\subsection{Problem Formulation and Design Overview}
\label{sec:s2s_problem_and_overview}

In sensing deployments involving sensor expansion or migration to a related structure, vibration measurements may be available at only a subset of the desired locations. We consider a reference sensor at position $a_{xyz}\in\mathbb{R}^{3}$ and an unmeasured target position $b_{xyz}\in\mathbb{R}^{3}$, defined in a common structural coordinate system. Given a reference vibration signal $x_{ref}\in\mathbb{R}^{L}$ containing $L$ temporal samples and synchronized RGB video $V\in\mathbb{R}^{T_v\times 3\times H\times W}$ containing $T_v$ frames, our objective is to reconstruct the response that would be observed at $b_{xyz}$. The target response is represented by an STFT spectrogram $Y_b\in\mathbb{R}^{F\times T_s}$, where $F$ and $T_s$ denote its frequency and temporal dimensions, respectively.

The central idea of \system{} is to infer the response at a new location or environment based on a vibration that has already been measured. The reference signal $x_{ref}$ reflects the combined effects of the mechanical excitation and structural dynamics present during an interaction. Its response at $b_{xyz}$, however, may differ substantially due to the propagation pathways and local mechanical conditions associated with the target position. Recovering the underlying forces and material parameters separately would require information unavailable in typical deployments. We instead use the measured vibration as the basis for reconstruction and exploit visual observations to characterize the structural and interaction-related factors that influence spatial transfer. Formally, the reconstruction is expressed as
\begin{equation}
    \hat{Y}_b = f_{\theta}(x_{ref}, a_{xyz}, b_{xyz}, V),
    \label{eq:s2s_target}
\end{equation}
where $\hat{Y}_b$ denotes the predicted target spectrogram and $f_{\theta}$ is the learned reconstruction function. During training, synchronized sensor pairs provide the measured target responses needed for supervision. Once trained, \system{} requires only the reference vibration, the two sensor coordinates, and the video to estimate the response at an unmeasured position.

\begin{figure}[!t]
    \centering
    \includegraphics[width=\linewidth]{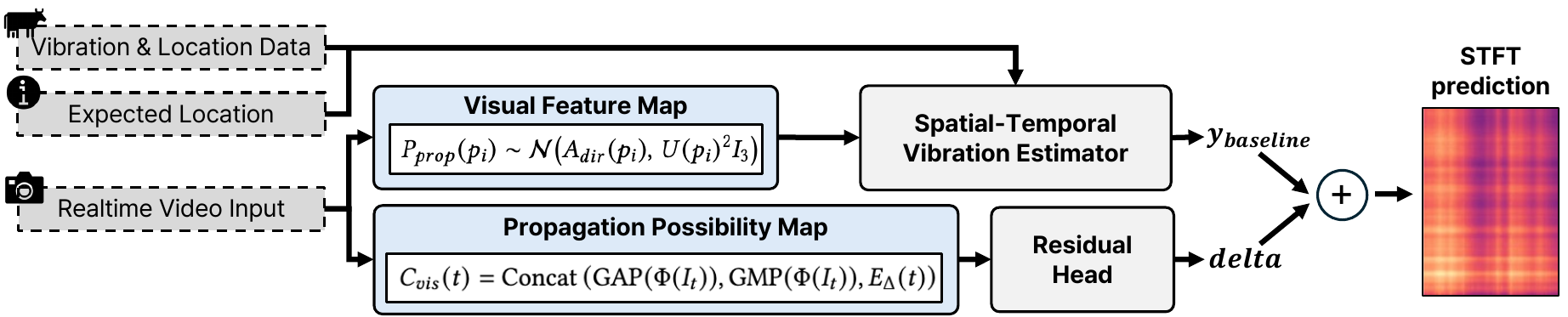}
    \caption{Overall architecture of \system{}.}
    \label{fig:framework_overview}
\end{figure}

Figure~\ref{fig:framework_overview} illustrates the overall architecture of \system{}. The \emph{Propagation Possibility Map} models structural transmission constraints by deriving a spatial pathway distribution $P_{prop}$ and its associated uncertainty $U$ from the video input and spatial geometry. Complementing structural pathways, the \emph{Visual Feature Map} captures source kinetics and surface interactions into a temporal visual representation $C_{vis}(t)$. Using these representations, the model executes a residual estimation pipeline, where the \emph{Spatial-Temporal Vibration Estimator} integrates $P_{prop}$ with the reference vibration $x_{ref}$ and sensor locations $(a_{xyz}, b_{xyz})$ to generate a baseline prediction $y_{baseline}$, while the \emph{Residual Head} derives a fine-grained offset $\delta$ from $C_{vis}(t)$. The two predictions are combined ($y_{baseline} + \delta$) to construct the complete STFT prediction $\hat{Y}_b$.

\subsection{Vision-Guided Physical Context Representation}
\label{sec:s2s_visual_context}

Estimating vibration at an unmeasured sensor position requires accounting for both the medium through which the vibration propagates and the interaction that excites it. These factors affect the measured response in different ways. Structural geometry provides relatively stable evidence about possible transmission pathways, while movements and surface appearance offer clues about the excitation and contact conditions that vary across observations. Although a monocular video cannot fully determine either process, it provides useful complementary information for spatial reconstruction. We therefore represent the physical context through a Propagation Possibility Map $P_{prop}$ with geometric uncertainty $U$, and a temporal visual representation $C_{vis}(t)$ describing interaction-related changes and appearance. Both representations guide the interpretation of a real reference measurement at another sensor position.

\subsubsection{Vision-Derived Propagation Possibility Map}
\label{sec:s2s_propagation}

Structure-borne vibration depends on the physical pathways connecting excitation sources to receiving sensors. For example, a continuous pipe provides a plausible transmission route along its length, whereas bends and junctions alter the geometry encountered by propagating waves. Structural continuity and orientation therefore provide useful evidence about where vibrations may travel. However, reconstructing these pathways from a monocular view is inherently uncertain, as joints, supports, and hidden connections may not be visible. To exploit the available geometry without requiring a complete structural model, we introduce the \emph{Vision-Derived Propagation Possibility Map} $P_{prop}$, which describes plausible transmission pathways together with a geometric uncertainty map $U$ indicating the reliability of the inferred structural information.

To characterize these pathways from visual observations, we first recover the spatial geometry of the underlying structure. A two-dimensional image provides limited information about physical connectivity, as structural elements at different depths may appear adjacent or overlap due to perspective. To address this limitation, we employ MoGe-2~\cite{wang2025moge2} to estimate three-dimensional scene points from a single RGB frame of the synchronized video $V$, yielding a point set $P$. Given that structural orientation and continuity can vary across different regions, we analyze the reconstructed geometry locally rather than assuming a uniform structural direction.

In particular, for each reconstructed point $p_i\in P$, we construct a local neighborhood $\mathcal{N}_i$ to characterize how nearby structural points are spatially distributed. Since more distant points may belong to different structural elements, particularly around bends and junctions, we assign distance-decaying weights $w_{ij}$ to emphasize the geometry immediately surrounding each point:
\begin{equation}
\begin{aligned}
    \mathcal{N}_i = \{p_j\in P:\|p_j-p_i\|_2\leq r\}, \quad w_{ij} = \exp\left(-\frac{\|p_j-p_i\|_2^2}{2\sigma^2}\right),
    \quad p_j\in\mathcal{N}_i,
\end{aligned}
\label{eq:s2s_local_neighborhood}
\end{equation}
where $r$ defines the neighborhood radius and $\sigma$ controls the spatial weighting. The resulting weighted distribution within $\mathcal{N}_i$ reflects how the structure extends around $p_i$, providing a basis for estimating its local orientation.

To capture these characteristics, we employ distance-weighted principal component analysis (PCA), which identifies the dominant direction along which the neighboring points are distributed. For instance, points on an elongated pipe or beam tend to extend farther along its length than across its width. By identifying the direction of greatest spatial spread, PCA allows us to approximate the orientation of the underlying structural element without requiring a predefined geometric model. We denote this direction at each point $p_i$ as $A_{dir}(p_i)$, providing a geometric cue about potential vibration transmission pathways. PCA also produces three eigenvalues, $\lambda_{i,1}\geq\lambda_{i,2}\geq\lambda_{i,3}$, which describe the spread of neighboring points along the three principal axes. These values further characterize the local point distribution and are used in the subsequent uncertainty estimation. For the final spatial integration, these three eigenvalues are averaged to form a single scalar summary representing the overall structural spread.

However, the direction $A_{dir}(p_i)$ inferred from PCA may not always reflect the actual structure, particularly when the reconstructed geometry is incomplete or inaccurate. For instance, insufficient neighboring points or inaccurate depth estimates can distort their spatial distribution, causing PCA to identify a misleading direction. Moreover, joints and connections hidden from the camera may still transmit vibrations despite being absent from the reconstructed geometry. These limitations suggest that the inferred structural directions should be considered according to the reliability of the visual observations. 

To account for such uncertainty, we estimate how reliably the reconstructed geometry supports the inferred direction $A_{dir}(p_i)$. Intuitively, a direction estimated from a dense and consistently distributed set of neighboring points is more trustworthy than one inferred from sparse or irregular observations. Similarly, inconsistencies in surface orientation or depth estimates may indicate that the reconstructed geometry does not accurately represent the physical structure. Based on these observations, we estimate a geometric uncertainty value $U(p_i)$ using four complementary indicators. To this end, $u_{neighbors}$ captures the availability of neighboring points within $\mathcal{N}_i$, while $u_{scatter}$ describes the relative spatial spread of these points along the principal axes identified by PCA. Meanwhile, $u_{normal}$ and $u_{depth}$ reflect inconsistencies in neighboring surface normals and estimated depth, respectively. Combining their normalized values, we define
\begin{equation}
    U(p_i)=\operatorname{Mean}\!\left(
    u_{neighbors}(p_i),u_{scatter}(p_i),
    u_{normal}(p_i),u_{depth}(p_i)
    \right).
    \label{eq:s2s_uncertainty}
\end{equation}
Here, each indicator quantifies a specific geometric risk factor derived from the local point distribution. First, $u_{neighbors}(p_i)$ penalizes regions where the available neighbor count falls below a minimum required threshold (threshold is set to 8 based on human prior). Second, $u_{scatter}(p_i)=\lambda_{i,3}/\max(\lambda_{i,1},\epsilon)$ describes the relative spatial spread along the smallest and largest principal axes obtained from PCA, where $\epsilon$ prevents division by zero. Third, $u_{normal}(p_i)$ measures the misalignment between the reconstructed surface normals and the inferred structural tensor direction. Finally, $u_{depth}(p_i)$ reflects the relative inconsistency of the local depth estimate compared to a 95th-percentile reference depth scale. Together, these indicators provide an estimate of geometric uncertainty, allowing the model to distinguish regions with reliable structural evidence from those where the inferred direction may be less trustworthy. Alongside this uncertainty, we also compute a geometric continuity cue $q$, which evaluates the local structural integrity by combining point density, surface normal alignment, and depth consistency into a single confidence score. The complete step-by-step procedure for extracting the local structural direction and its associated geometric uncertainty is summarized in Algorithm~\ref{alg:pca_uncertainty}.

\begin{algorithm}[h]
\caption{Local PCA and Geometric Uncertainty Computation}
\label{alg:pca_uncertainty}
\begin{algorithmic}[1]
\Require 3D point set $P$, neighborhood radius $r$, smoothing factor $\sigma$
\For{each point $p_i\in P$}
    \State $\mathcal{N}_i\leftarrow\{p_j\in P:\lVert p_j-p_i\rVert_2\leq r\}$
    \State $w_{ij}\leftarrow\exp(-\lVert p_j-p_i\rVert_2^2/(2\sigma^2))$, $p_j\in\mathcal{N}_i$
    \State $\mu_i\leftarrow\sum_{j\in\mathcal{N}_i}w_{ij}p_j/\sum_{j\in\mathcal{N}_i}w_{ij}$
    \State $C_i\leftarrow\sum_{j\in\mathcal{N}_i}w_{ij}(p_j-\mu_i)(p_j-\mu_i)^{\mathsf T}/\sum_{j\in\mathcal{N}_i}w_{ij}$
    \State $\lambda_i,v_i\leftarrow\operatorname{Eigh}(C_i)$ \Comment{Sort eigenvalues in descending order}
    \State $A_{dir}(p_i)\leftarrow v_{i,1}$ \Comment{Dominant geometric direction}
    \State $u_{scatter}(p_i)\leftarrow\lambda_{i,3}/\max(\lambda_{i,1},\epsilon)$
    \State $U(p_i)\leftarrow\operatorname{Mean}(u_{neighbors},u_{scatter},u_{normal},u_{depth})$
\EndFor
\end{algorithmic}
\end{algorithm}

Having estimated the dominant structural direction $A_{dir}(p_i)$ and its geometric uncertainty $U(p_i)$, we next consider how these two quantities can jointly represent potential vibration propagation pathways. While $A_{dir}(p_i)$ indicates the direction along which the visible structure extends, $U(p_i)$ provides an indication of how reliably this direction can be inferred. Accordingly, the estimated direction should provide stronger guidance when the geometry is reliably reconstructed, while allowing greater flexibility when observations are incomplete or uncertain. For instance, a clearly observed pipe provides strong evidence for propagation along its length, whereas an occluded junction may contain additional transmission pathways that cannot be inferred from the visible geometry alone. 

To capture this relationship, we formulate the propagation possibility map $P_{prop}$ as a probabilistic directional representation whose spread depends on the estimated geometric uncertainty:
\begin{equation}
    \mathbf{s}_{prop}(p_i) = \left[ q(p_i), U(p_i), \bar{\lambda}_i \right]^T
    \label{eq:s2s_propagation_prior}
\end{equation}
where $A_{dir}(p_i)$ represents the preferred local orientation, $U(p_i)$ controls the distribution's spread, and $I_3$ is the three-dimensional identity matrix. This formulation allows the propagation map $P_{prop}$ to emphasize directions supported by reliable structural observations while retaining flexibility in geometrically uncertain regions. In practice, the neural network does not perform explicit probability computations or randomly sample directions from this distribution. Instead, the core parameters of the distribution, specifically a scalar summary of the structural tensor eigenvalues, a geometric continuity cue $q$, and the uncertainty $U(p_i)$, are extracted as deterministic features. These parameters are then projected as multi-channel feature maps onto a discrete spatial grid, thereby guiding vibration reconstruction without assuming complete knowledge of the underlying propagation pathways.

Overall, \system{} uses the Vision-Derived Propagation Possibility Map $P_{prop}$ to integrate reconstructed 3D geometry $P$ by projecting the extracted structural parameters as feature channels onto a shared spatial grid. Specifically, each local spatial bin extracts a 3-dimensional descriptor composed of the geometric continuity cue, geometric uncertainty, and the eigenvalue summary, defined as $[q, U, \bar{\lambda}]$. Across $K$ discretized spatial neighborhood bins along the wave propagation path ($K=8$ in our deployment), this yields a total of $3K = 24$ dedicated wave propagation channels. This channel-wise integration provides deterministic, physically grounded spatial guidance for vibration reconstruction between reference and target sensor positions $(a_{xyz},b_{xyz})$.

\subsubsection{Visual Feature Map for Source Kinetics and Implicit Material Dynamics}
\label{sec:s2s_source_material}

The propagation possibility map $P_{prop}$ describes plausible routes through which vibrations may travel, but a structural route alone cannot determine the response recorded by a sensor. Even along the same pipe, a brief impact and a sustained contact can excite different vibration patterns. The surfaces involved in the interaction may further affect how the vibration is generated and attenuated. Precisely measuring these contact forces and material properties would require additional instrumentation and detailed physical calibration. We therefore introduce the \emph{Visual Feature Map for Source Kinetics and Implicit Material Dynamics} to capture two kinds of visual evidence that complement $P_{prop}$, namely changes in visible activity that may indicate when a structure is excited, and appearance characteristics that provide context about the surfaces involved. We represent this information using a temporal visual feature $C_{vis}(t)$ derived from the synchronized video $V$.

We first consider how video can reveal changes in physical interactions that may excite the structure. When an animal steps against a metal rail, for example, the contact may be brief even though the resulting vibration persists for a longer period. Such interactions often involve visible movements, making temporal changes between consecutive frames useful cues for identifying potential excitation events. A common approach in computer vision is frame differencing, which detects changes by comparing consecutive images. However, differences in raw pixel values can also arise from illumination changes or other visual variations unrelated to physical interactions. To obtain more meaningful temporal information, we instead compare learned visual features, allowing the model to characterize changes in the observed scene beyond pixel-level differences.

To achieve this, we employ a visual encoder $\Phi$ to extract a high-level feature map $\Phi(I_t) \in \mathbb{R}^{C \times H' \times W'}$ from each RGB frame $I_t$ in the synchronized video $V$. To aggregate spatial information into channel-wise representations, we define global average pooling ($\operatorname{GAP}$) and global max pooling ($\operatorname{GMP}$) for any spatial feature tensor $X \in \mathbb{R}^{C \times H' \times W'}$ at channel $c \in \{1, \dots, C\}$ as:
\begin{equation}
    \operatorname{GAP}(X)_c = \frac{1}{H' W'} \sum_{h=1}^{H'} \sum_{w=1}^{W'} X_{c, h, w}, \quad
    \operatorname{GMP}(X)_c = \max_{h, w} X_{c, h, w}.
    \label{eq:s2s_pooling_defs}
\end{equation}
Using these operators, we first capture the dynamic interactions that potentially excite the structure. We compute the element-wise absolute difference between consecutive feature maps to suppress ambient illumination variations and isolate physical motion. Spatially aggregating these changes via $\operatorname{GAP}$ yields a channel-wise temporal descriptor $E_{\Delta}(t) \in \mathbb{R}^C$:
\begin{equation}
    E_{\Delta}(t) = \operatorname{GAP}\left( \left| \Phi(I_t) - \Phi(I_{t-1}) \right| \right).
    \label{eq:s2s_delta_energy}
\end{equation}
The resulting descriptor captures the onset and intensity of potentially vibration-inducing activities, whose mapping to mechanical excitations is supervised by synchronized vibration data.

While the temporal feature-difference descriptor $E_{\Delta}(t)$ captures changes in visible activity, it provides limited information about the surfaces involved in those interactions. For instance, a similar movement may produce different vibration responses depending on whether an animal contacts a rigid metal rail or a softer surface, due to differences in their mechanical properties. Although visual appearance cannot directly reveal stiffness or damping coefficients, it offers clues about the surface characteristics that may influence the resulting vibration. 

To incorporate such information, we additionally exploit the spatial feature map $\Phi(I_t)$ extracted by the visual encoder $\Phi$. In detail, we employ GAP again to summarize features distributed across the scene and global max pooling (GMP) to preserve prominent features from smaller regions that might otherwise be diluted by averaging. Combining these dual spatial pooling operations allows us to capture both broadly distributed and localized appearance characteristics without explicitly estimating material properties.

We combine the extracted appearance features with the temporal feature-difference descriptor $E_{\Delta}(t)$ to construct the temporal visual representation $C_{vis}(t)$:
\begin{equation}
    C_{vis}(t)=\operatorname{Concat}\left(
        \operatorname{GAP}(\Phi(I_t)),
        \operatorname{GMP}(\Phi(I_t)),
        E_{\Delta}(t)
    \right).
    \label{eq:s2s_visual_token}
\end{equation}
We implement the visual encoder $\Phi$ using ResNet-18 with its early layers frozen during training. Each descriptor contains 512 channels, yielding a 1,536-dimensional temporal visual representation $C_{vis}(t)$. 

Overall, \system{} uses the Visual Feature Map for Source Kinetics and Implicit Material Dynamics, represented by $C_{vis}(t)$, to complement the structural guidance provided by the Vision-Derived Propagation Possibility Map $P_{prop}$ with information about dynamic interactions and surrounding surface characteristics. The following module integrates these complementary representations with the measured reference vibration to reconstruct the target response.

\subsection{Spatial-Temporal Wave Propagation Architecture}
\label{sec:s2s_wave_estimator}

Two aforementioned visual representations provide complementary information about where vibrations may propagate and how physical interactions may influence the resulting response. However, these visual representations cannot fully capture the actual excitation forces and mechanical conditions, whose combined effects are already reflected in the measured reference vibration $x_{ref}$. Leveraging this observation, we focus on learning how the measured vibration changes between reference and target sensor positions $(a_{xyz},b_{xyz})$, guided by the structural and interaction-related information extracted from video. To achieve this, we develop the \emph{Spatial-Temporal Wave Propagation Architecture}, which integrates the reference vibration, sensor coordinates, and both visual representations through a residual formulation consisting of the \emph{Spatial-Temporal Vibration Estimator} and the \emph{Residual Head}.

We first represent the measured vibration as an ordered sequence of short segments. For instance, a brief impact against a metal pipe can produce an initial vibration followed by oscillations that persist after the contact, making the temporal characteristics of the measured signal important for reconstruction. To preserve these patterns, we encode each segment of the reference vibration $x_{ref}$ using a multilayer perceptron (MLP), producing a sequence of sensor tokens for subsequent temporal modeling.

While the reference vibration captures the actual structural response, reconstructing it at another position requires accounting for the spatial relationship between the two sensors. Even under the same excitation, the received vibration may differ depending on the sensor locations and their relative geometry. Thus, the estimator needs to consider both where the vibration was originally measured and where its response is to be reconstructed.

To incorporate this information, we use the reference sensor position $a_{xyz}$ and target sensor position $b_{xyz}$, together with their relative displacement $\Delta_{ab}=b_{xyz}-a_{xyz}$ and Euclidean distance $d_{ab}=\lVert\Delta_{ab}\rVert_2$. While the absolute coordinates identify the sensor locations, their displacement and distance describe the relative spatial relationship. Precisely, we construct a ten-dimensional sensor-pair descriptor $z_{ab}$ as
\begin{equation}
    z_{ab}=\operatorname{Concat}\left(
    a_{xyz},b_{xyz},\Delta_{ab},
    \log(d_{ab}+\epsilon)
    \right),
    \label{eq:s2s_coordinate}
\end{equation}
where $\epsilon>0$ ensures numerical stability, and the logarithmic term compresses the range of spatial separation.

Although the sensor-pair descriptor $z_{ab}$ captures the relative geometry of the two sensors, it does not fully characterize the structural pathways connecting them. For instance, two sensor pairs may have similar spatial distances, while one lies along a continuous metal pipe and the other spans a junction between structural elements. Their vibration responses may therefore differ due to distinct propagation conditions.

To account for these structural transmission factors, the Spatial-Temporal Vibration Estimator incorporates the Vision-Derived Propagation Possibility Map $P_{prop}$, which provides spatial guidance about plausible propagation pathways. Here, the propagation map $P_{prop}$ is spatially aligned to the specific sensor pair via nearest-neighbor coordinate matching. We project $P_{prop}$ and the sensor-pair descriptor $z_{ab}$ into compatible feature dimensions and align them with the reference sensor tokens.

However, the reliability of structural cues may vary across locations. A clearly observed pipe may provide useful directional guidance, whereas an occluded junction can contain hidden connections that are not reflected in the reconstructed geometry. Assigning equal importance to all contextual features could therefore introduce misleading information during reconstruction.

To address this limitation, we employ gated additive fusion to adaptively regulate the contributions of the sensor-pair and structural features. In this scheme, learnable gates modulate the projected features before adding them to the encoded reference vibration. We inject $U$ as an explicit physical feature channel alongside the propagation representations. This implicit, data-driven formulation allows the network to learn how to adaptively adjust its reliance on visual context while preserving the measured vibration as the basis for reconstruction.

Vibration responses evolve over time as waves propagate, reflect at structural boundaries, and gradually decay. For example, a brief impact on a metal rail may influence the measured vibration across multiple consecutive segments, making independent segment-wise predictions insufficient to capture the resulting temporal dependencies. We therefore employ a bidirectional long short-term memory (BiLSTM) network to process the fused sensor representations, producing a baseline prediction $y_{baseline}$.

Concurrently, local contact dynamics and surface interactions can induce high-frequency variations that are not fully captured by structural propagation alone. To model these localized deviations, the \emph{Residual Head} processes the temporal visual representation $C_{vis}(t)$ to compute a fine-grained residual offset $\delta$. The final target STFT spectrogram prediction is then obtained by combining both branches:
\begin{equation}
    \hat{Y}_b = y_{baseline} + \delta.
    \label{eq:s2s_residual_sum}
\end{equation}

Overall, the Spatial-Temporal Wave Propagation Architecture integrates the measured reference vibration, sensor-pair geometry, and vision-derived physical context to generate vibration data for unmeasured sensor locations.

\subsection{Physics-Informed Training Criteria}
\label{sec:s2s_training}

We optimize \system{} using a decoupled two-stage training scheme that aligns global spectral representations with fine-grained elastodynamic residual corrections. Rather than relying on a single end-to-end objective, each stage employs a dedicated loss function tailored to its specific structural representation role.

\textbf{Stage 1: Baseline Teacher Objective.} In the first stage, the baseline model is trained using the measured target spectrogram $Y_b$ as ground truth to capture the primary spectral trajectory and spatial wave propagation across sensor locations, optimizing the baseline prediction $y_{baseline}$. However, standard point-wise regression objectives tend to smooth over short-lived impact events. To preserve both broadband energy distributions and localized shock transients, we formulate a Composite Regression Loss $\mathcal{L}_{\text{Stage-1}}$:
\begin{equation}
    \mathcal{L}_{\text{Stage-1}} = \mathcal{L}_{\text{MR-STFT}}(Y_b, y_{baseline}) + \alpha_1 \mathcal{L}_{\text{grad}} + \alpha_2 \mathcal{L}_{\text{peak}} + \alpha_3 \mathcal{L}_{\text{margin}},
    \label{eq:stage1_loss}
\end{equation}
where $\mathcal{L}_{\text{MR-STFT}}$ evaluates spectral consistency across multiple time-frequency resolutions, $\mathcal{L}_{\text{grad}}$ penalizes high-frequency boundary smoothing via spectrogram gradients, $\mathcal{L}_{\text{peak}}$ applies an asymmetric penalty to prevent underestimating sharp transient impacts, and $\mathcal{L}_{\text{margin}}$ enforces representation contrast against background noise.

\textbf{Stage 2: Residual Adaptation Objective.} In the second stage, the trained Stage~1 baseline weights are frozen, and the residual head is introduced to predict the high-frequency physical offset $\delta$ from the temporal visual representation $C_{vis}(t)$, yielding the final output $\hat{Y}_b = y_{baseline} + \delta$. The residual objective focuses on fine-grained wavefield alignment and regularization:
\begin{equation}
    \mathcal{L}_{\text{Stage-2}} = \mathcal{L}_{\text{MSE}}(Y_b, \hat{Y}_b) + \lambda_{\text{L1}} \mathcal{L}_{\text{L1}}(Y_b, \hat{Y}_b) + \lambda_{\delta} \|\delta\|_2^2,
    \label{eq:stage2_loss}
\end{equation}
where the combined MSE and L1 terms enforce precise signal-level and envelope fidelity, while the $L_2$ regularization penalty $\|\delta\|_2^2$ prevents the residual head from drifting into ungrounded high-amplitude artifacts.

Separating these criteria into two stages ensures that the base model learns stable structural manifolds first, after which the residual network refines localized boundary-induced elastodynamic shifts without catastrophic interference.

\section{Evaluation}
\label{sec:Evaluation}

Building upon the limitations identified in the preliminary study, we evaluate our proposed physics-informed cross-modal architecture. The goal is to demonstrate that by grounding the vibration data with visual physical context, our model can successfully predict vibration signals at arbitrary, unseen target locations.

\subsection{Dataset \& Learning Phase Explanation}
\label{subsec:dataset_explanation}

The dataset utilized in this work was collected from an operational commercial pig farm by deploying vibration sensor (Geophone~\cite{sparkfun_geophone_sm24}) alongside time-synchronized video recording systems. 

\begin{table}[htbp]
  \centering
  \caption{Crate Group Duration Summary}
  \label{tab:crate_duration}
  \begin{tabular}{llrrr}
    \toprule
    \textbf{Crate} & \textbf{Crate Group} & \textbf{Seen Loc (2,3,5) [hours]} & \textbf{Unseen Loc (1,4) [hours]} & \textbf{Total [hours]} \\
    \midrule
    Crate\_02 & Seen   & 139.692 &   0.000 & 139.692 \\
    Crate\_03 & Seen   & 139.827 &   0.000 & 139.827 \\
    Crate\_08 & Seen   & 277.185 & 185.233 & 462.419 \\
    Crate\_04 & Unseen &  68.455 &   0.000 &  68.455 \\
    Crate\_06 & Unseen &  69.896 &   0.000 &  69.896 \\
    Crate\_07 & Unseen & 214.073 & 142.661 & 356.734 \\
    \bottomrule
  \end{tabular}
\end{table}

As detailed in Table~\ref{tab:crate_duration}, data were collected across six distinct crates, comprising two structurally identical configurations with three crates per group. Regarding sensor placement, three sensor positions traversing the crate diagonally from the lower-left to the upper-right corner are designated as \textit{seen sensor locations} (Location 2, 3, and 5), whereas the remaining two positions are classified as \textit{unseen sensor locations} (Location 1 and 4). For Crates 02, 03, 04, and 06, geophone recordings were collected using only the three seen sensor locations. 

During the model training phase, only Crates~02 and 03 within the seen crate and seen sensor location category (\texttt{seen\_train}) are used for optimization. Although Crate~08 belongs to the same structural category and contains seen sensor locations, it is strictly held out during training and reserved exclusively for evaluation (\texttt{seen\_eval}).

All experiments, model training, and evaluations were conducted on an NVIDIA GeForce RTX 5070~Ti GPU~\cite{nvidia_rtx_5070_ti}. The learning pipeline follows a two-stage residual adaptation scheme focusing on vibration wave propagation characteristics:

\begin{enumerate}
    \item \textbf{Stage 1 (Baseline Teacher Training):} A physics-free baseline model is first trained on the vision modality to establish an initial predictive teacher checkpoint. The feature extractor and localization estimator networks from this stage capture base visual representations and provide baseline coordinate predictions.
    
    \item \textbf{Stage 2 (Propagation-Only Residual Learning):} In this stage, the pre-trained Stage~1 teacher weights are frozen, and a physics-guided residual framework is introduced. The complete physical feature space comprises 24 wave propagation channels. Only the physics encoder and the residual correction head are updated, predicting an additive offset $\delta$ such that the final output is defined as $\hat{y} = y_{\text{baseline}} + \delta$.
    
    \item \textbf{Optimization \& Objective Function:} The learning process utilizes distinct objective functions for each stage. For Stage 1, the baseline teacher is optimized using the Composite Regression Loss (as detailed in Section \ref{sec:s2s_training}), encompassing the multi-resolution spectral, gradient, peak-weighted, and contrastive margin losses. For Stage 2, the residual network is optimized using AdamW with an initial learning rate of $\eta = 1\times 10^{-4}$ and weight decay of $1\times 10^{-4}$. Automatic Mixed Precision (AMP, FP16) and gradient clipping with a maximum norm threshold of $1.0$ are applied for numerical stability. The Stage 2 objective function balances primary trajectory reconstruction and residual regularization:
    \begin{equation}
        \mathcal{L}_{\text{Stage-2}} = \mathcal{L}_{\text{MSE}}(y, \hat{y}) + \lambda_{\text{L1}} \mathcal{L}_{\text{L1}}(y, \hat{y}) + \lambda_{\delta} \|\delta\|_2^2
    \end{equation}
    where $\lambda_{\text{L1}} = 0.2$ and $\lambda_{\delta} = 1\times 10^{-4}$.
    
    \item \textbf{Training Schedule \& Evaluation Protocol:} Under the standard \texttt{seen\_baseline} protocol, the model is trained with a batch size of 16 using 2 DataLoader worker threads for up to 100 epochs, monitored by early stopping with a patience of 3 epochs based on the validation Mean Squared Error (MSE). The checkpoint achieving the lowest validation MSE is selected and subsequently evaluated across all defined seen and unseen benchmark phases.
\end{enumerate}

\subsection{Overall performance}
We evaluate the performance of \system{} by first conducting an end-to-end evaluation that demonstrates the primary objective of the model.
To this end, we quantitatively and qualitatively analyze whether our proposed approach effectively reconstructs waveforms while adhering to physical laws.
Next, to examine whether the generated waveforms contain features that are practically learnable, we train a classifier on the data generated by each model and evaluate its performance on a downstream task.

\subsubsection{Waveform Reconstruction Performance}
The primary objective of our system is to accurately synthesize target vibration responses in the form of short-time Fourier transform (STFT) spectrograms, faithfully replicating the true physical wavefield observed by contact sensors. For evaluation, we utilized 200 hours of unseen data (50 hours per environment) on which the model was not trained. To quantitatively evaluate the fidelity of the reconstructed time-frequency energy distribution against ground truth measurements, we employ two complementary evaluation metrics: Mean Squared Error (MSE) and Mean Absolute Error (MAE).

In structure-borne elastodynamics, these two metrics capture distinct physical aspects of wave propagation:
\begin{itemize}
    \item \textbf{Mean Squared Error (MSE):} When a model fails to capture sudden mechanical impact shocks (e.g., transient contact events) or dramatically misestimates dominant resonant energy peaks, the metric surges sharply. Because MSE penalizes large discrepancies quadratically, it imposes penalties on localized errors occurring in transient, high-energy impact bursts and dominant modal frequencies. Thus, MSE serves as a sensitive indicator of whether the model reliably reconstructs critical peak impact dynamics without suffering from amplitude misestimations.
    \item \textbf{Mean Absolute Error (MAE):} When evaluating the broader consistency of ambient structural decay profiles, background vibration levels, and wideband signal attenuation, MAE reflects whether the overall spectral energy scale is well-maintained over time. By averaging absolute errors uniformly across all time-frequency bins, it remains resilient to isolated extreme outliers and broadband measurement noise, providing a stable measure of global baseline conformity across the full spectrogram.
\end{itemize}

\begin{figure*}[t]
\centering

\begin{minipage}{\linewidth}
  \centering
  \captionof{table}{Reconstruction performance (MSE and MAE) across different crate and sensor location settings\\(Loc. = Sensor Location).}
  \label{tab:reconstruction_results}
  \label{tlr}
  \resizebox{\linewidth}{!}{%
  \begin{tabular}{lcccccccc}
  \toprule
  \multirow{2}{*}{\textbf{Method}} & \multicolumn{2}{c}{\textbf{Seen Crate + Seen Loc.}} & \multicolumn{2}{c}{\textbf{Seen Crate + Unseen Loc.}} & \multicolumn{2}{c}{\textbf{Unseen Crate + Seen Loc.}} & \multicolumn{2}{c}{\textbf{Unseen Crate + Unseen Loc.}} \\
  \cmidrule(lr){2-3} \cmidrule(lr){4-5} \cmidrule(lr){6-7} \cmidrule(lr){8-9}
  & MSE $\downarrow$ & MAE $\downarrow$ & MSE $\downarrow$ & MAE $\downarrow$ & MSE $\downarrow$ & MAE $\downarrow$ & MSE $\downarrow$ & MAE $\downarrow$ \\
  \midrule
  Naive & 2.0276  & 1.1441 & 77.5576 & 3.7369 & 2.3286  & 1.2271 & 87.3275 & 4.8497 \\
  Audio & 3.2567  & 1.5863 & 1.4235  & 0.9359 & 1.7861  & 1.0000 & 3.2683  & 1.5556 \\
  FEM   & 97.5851 & 7.8023 & 98.2487 & 9.8259 & 93.6060 & 9.5800 & 98.1402 & 7.8084 \\
  Ours  & 0.8605  & 0.7170 & 1.2650  & 0.8697 & 0.9137  & 0.7949 & 1.0212  & 0.7883 \\
  \bottomrule
  \end{tabular}%
  }
\end{minipage}

\vspace{1.2em} 

\begin{minipage}{\linewidth}
  \centering
  \includegraphics[width=\linewidth]{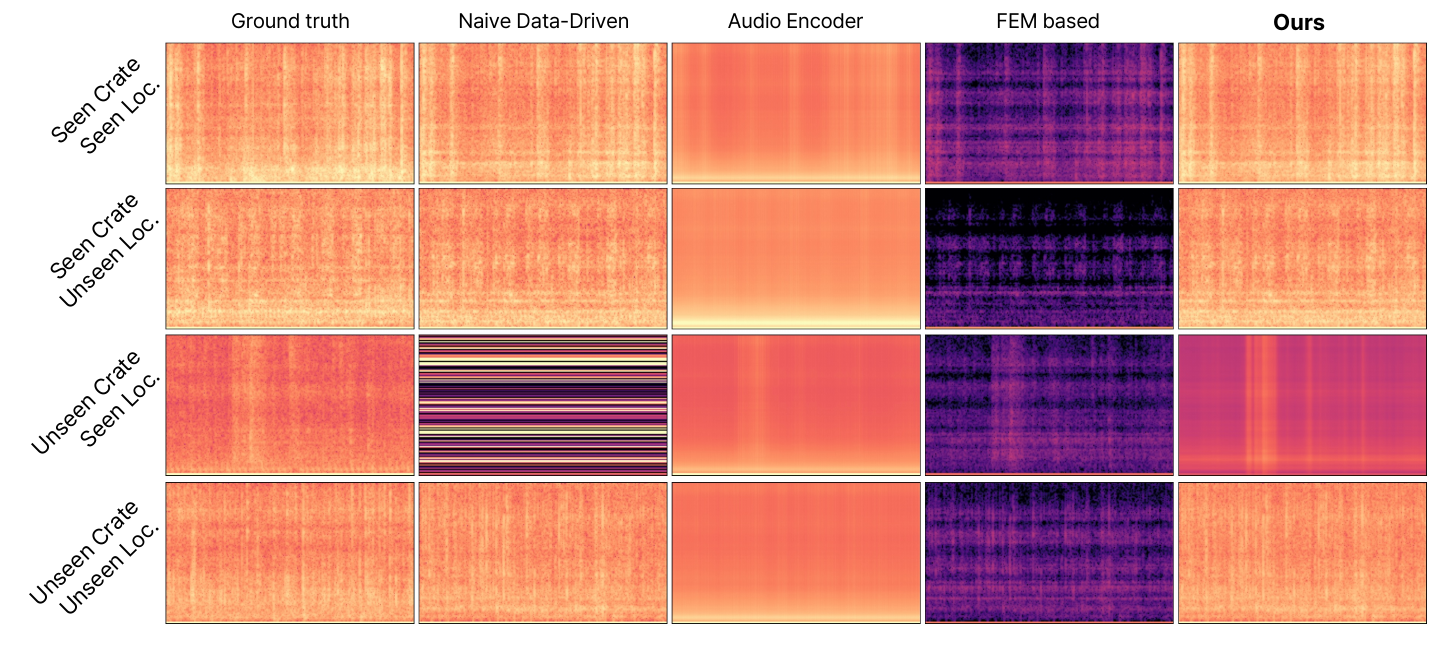}
  \captionof{figure}{Qualitative analysis on synthetic stft data by method (Loc. = Sensor Location)}
  \label{fig:qualitative_analysis_stft}
\end{minipage}

\end{figure*}

As shown in Table~\ref{tab:reconstruction_results}, \system{} consistently outperforms all baseline approaches across both seen and unseen spatial transfer scenarios in terms of both MSE and MAE. In the in-domain baseline setting (\textit{Seen Crate + Seen Sensor Loc.}), \system{} achieves an MSE of 0.8605 and an MAE of 0.7170, reducing the reconstruction error by more than 57.5\% in MSE compared to the purely data-driven sequence-to-sequence baseline (Naive: 2.0276). 

Crucially, the limitations of conventional paradigms become pronounced when subjected to spatial and structural domain shifts:

First, as qualitatively observed in Figure~\ref{fig:qualitative_analysis_stft}, the \textbf{Naive Data-Driven} approach appears to superficially mimic the ambient environment under familiar conditions. However, under spatial and structural transfers, most conspicuously in the \textit{Unseen Crate + Seen Sensor Loc.} setting, it suffers from generative collapse, producing meaningless, degraded striping patterns across all frequencies. Such failure typically manifests when a purely data-driven sequence model severely overfits to its training domain (\textit{Seen Crate + Seen Sensor Loc.}) without learning underlying physical wave dynamics, rendering it incapable of flexibly generalizing to altered boundary conditions
. Consequently, as shown in Table~\ref{tab:reconstruction_results}, this over-reliance yields massive MSE (up to 87.3275) and MAE (up to 4.8497) errors under unseen configurations.

Second, while the \textbf{Audio Encoder} baseline benefits from massive acoustic pretraining to achieve comparatively lower MAE than the Naive model, it remains fundamentally mismatched to structural elastodynamics. In contrast to airborne acoustic environments, structure-borne vibration is characterized by dense background structural reverberations, induced by continuous animal movements and unmeasured micro-impact reflections, resulting in fine-grained, broadband frequency distributions rather than sparse harmonic peaks. As evident in Figure~\ref{fig:qualitative_analysis_stft}, the audio model over-simplifies the STFT background into an excessively smooth profile. Furthermore, as seen in the \textit{Seen Crate + Unseen Sensor Loc.} regime, it completely fails to resolve essential recurring transient patterns across time.

Third, although the \textbf{Physics-Based FEM} baseline captures the existence of oscillatory features, it suffers from a pronounced sim-to-real disparity. Because exact internal material properties, boundary damping ratios, and structural contact damping are fundamentally uncalibrated, FEM cannot accurately reproduce realistic attenuation rates or authentic boundary wave reflections. In Figure~\ref{fig:qualitative_analysis_stft}, this substantial energy mismatch is visually apparent from its substantially darker colormap, representing significantly underestimated amplitudes. More fundamentally, an idealized FEM model inherently vibrates at mathematically discrete eigenfrequencies governed rigidly by its nominal geometry, stiffness, and idealized support boundaries. When dynamic forces excite this closed geometry, waves reflect repeatedly within the structural boundaries, establishing dominant standing waves at specific modal frequencies (e.g., around bins 35, 50, 75, and 85) that persist horizontally across the entire time axis. These rigid horizontal resonance bands fail to reflect the non-linear, multi-contact, and broadband dampening characteristics of real-world deployments. Because real structural waves exhibit rapid phase dispersion and complex decay rather than stationary single-mode resonance, these persistent modal artifacts produce point-to-point discrepancy across every time-frequency bin, leading to excessively high MSE ($>93$) and MAE ($>7.8$) values across all evaluation regimes.

In contrast, as shown in Table~\ref{tab:reconstruction_results}, \system{} consistently outperforms all baseline approaches across both seen and unseen spatial transfer scenarios in terms of both MSE and MAE. In the in-domain baseline setting (\textit{Seen Crate + Seen Sensor Loc.}), \system{} achieves an MSE of 0.8605 and an MAE of 0.7170, reducing the reconstruction error by more than 57.5\% in MSE compared to the purely data-driven sequence-to-sequence baseline (Naive: 2.0276). This consistent quantitative superiority demonstrates that our framework reliably infers physically grounded vibration responses. As corroborated by Figure~\ref{fig:qualitative_analysis_stft}, \system{} reconstructs stable waveforms while faithfully preserving critical transient impacts and fine-grained spectral distributions, even under challenging unseen deployment conditions.

\subsubsection{Downstream Utility}
\label{subsubsec:downstream_utility}
We now evaluate whether the synthesized data preserves meaningful semantic features beyond basic physical reconstruction. While low MSE and MAE indicate high signal-level precision, relying solely on point-wise metrics risks overlooking whether the model merely mimics dominant spectral peaks and background envelopes without retaining the subtle behavioral characteristics embedded in the signal. To rigorously assess this downstream utility, we train a Convolutional Neural Network (CNN) classifier on the synthesized spectrograms to recognize four distinct target classes: \textit{eating+standing}, \textit{lying}, \textit{sitting}, and \textit{standing}. 

The classification experiment is conducted using an unseen dataset of 560~hours (140~hours per environment) that was completely excluded during the synthesis model's training phase. We leverage approximately 100~hours of paired data from this unseen set to generate synthetic vibration responses targeted at our desired sensor locations. Using this synthesized dataset, we segment the samples into 30-second windows with a 20-second step size (i.e., a 10-second overlap) to train the CNN classifier. To emulate a realistic edge deployment scenario, the trained classifier is subsequently evaluated on the remaining 40~hours using actual physical sensor measurements as input, thereby directly benchmarking how well representations learned from purely synthesized data transfer to real-world inference.

\begin{table}[h]
\centering
\caption{Downstream task performance evaluation across different crate and sensor location settings (Loc. = Sensor Location).}
\label{tab:downstream_results}
\resizebox{\linewidth}{!}{%
\begin{tabular}{lcccccccc}
\toprule 
\multirow{2}{*}{\textbf{Method}} & \multicolumn{2}{c}{\textbf{Seen Crate + Seen Loc.}} & \multicolumn{2}{c}{\textbf{Seen Crate + Unseen Loc.}} & \multicolumn{2}{c}{\textbf{Unseen Crate + Seen Loc.}} & \multicolumn{2}{c}{\textbf{Unseen Crate + Unseen Loc.}} \\
\cmidrule(lr){2-3} \cmidrule(lr){4-5} \cmidrule(lr){6-7} \cmidrule(lr){8-9}
& Acc. (\%) $\uparrow$ & Micro-F1 $\uparrow$ & Acc. (\%) $\uparrow$ & Micro-F1 $\uparrow$ & Acc. (\%) $\uparrow$ & Micro-F1 $\uparrow$ & Acc. (\%) $\uparrow$ & Micro-F1 $\uparrow$ \\
\midrule
Naive & 38.05\% & 0.3470 & 28.35\% & 0.1909 & 42.20\% & 0.2554 & 21.94\% & 0.1916 \\
Audio & 59.23\% & 0.4243 & 61.73\% & 0.4288 & 69.91\% & 0.2075 & 63.07\% & 0.1923 \\
FEM   & 0\% & 0.0 & 0\% & 0.0 & 0\% & 0.0 & 0\% & 0.0 \\
Ours  & 81.69\% & 0.8180 & 78.54\% & 0.7437 & 72.16\% & 0.6852 & 71.69\% & 0.6749 \\
\bottomrule
\end{tabular}%
} 
\end{table}

As shown in Table~\ref{tab:downstream_results}, our proposed model achieves the highest classification accuracy across all baselines while maintaining stability across every evaluated environment. 
In contrast, the Naive baseline exhibits its peak performance under the familiar in-distribution setting (\textit{Seen Crate + Seen Sensor Loc.}); however, as it encounters more generalized scenarios, both its classification accuracy and Micro-F1 score degrade notably. 
Notably, although Naive appears to achieve higher raw accuracy under \textit{Unseen Crate + Seen Sensor Loc.}, its Micro-F1 score collapses significantly. This disparity highlights an accuracy illusion driven by pathological class-collapse, where the model disproportionately predicts majority labels, a phenomenon directly aligned with the frequent output collapses observed in Figure~\ref{fig:qualitative_analysis_stft}. 
The Audio-based baseline presents a similar failure mode: while it sustains ostensibly competitive accuracy, its Micro-F1 score similarly breaks down. As detailed further in Section~\ref{subsec:system_robustness}, this discrepancy exemplifies how purely point-wise reconstruction metrics like MAE and MSE fail to capture whether essential discriminative patterns are preserved. 
Finally, the FEM baseline yields synthetic data whose underlying distribution critically deviates from physical reality, causing the classifier trained on its generated samples to fail entirely when transferred to the actual sensor measurement distribution.

\subsubsection{Adaptation to Finetuning}

\begin{figure}[htbp]
\centering
    \includegraphics[width=0.9\linewidth]{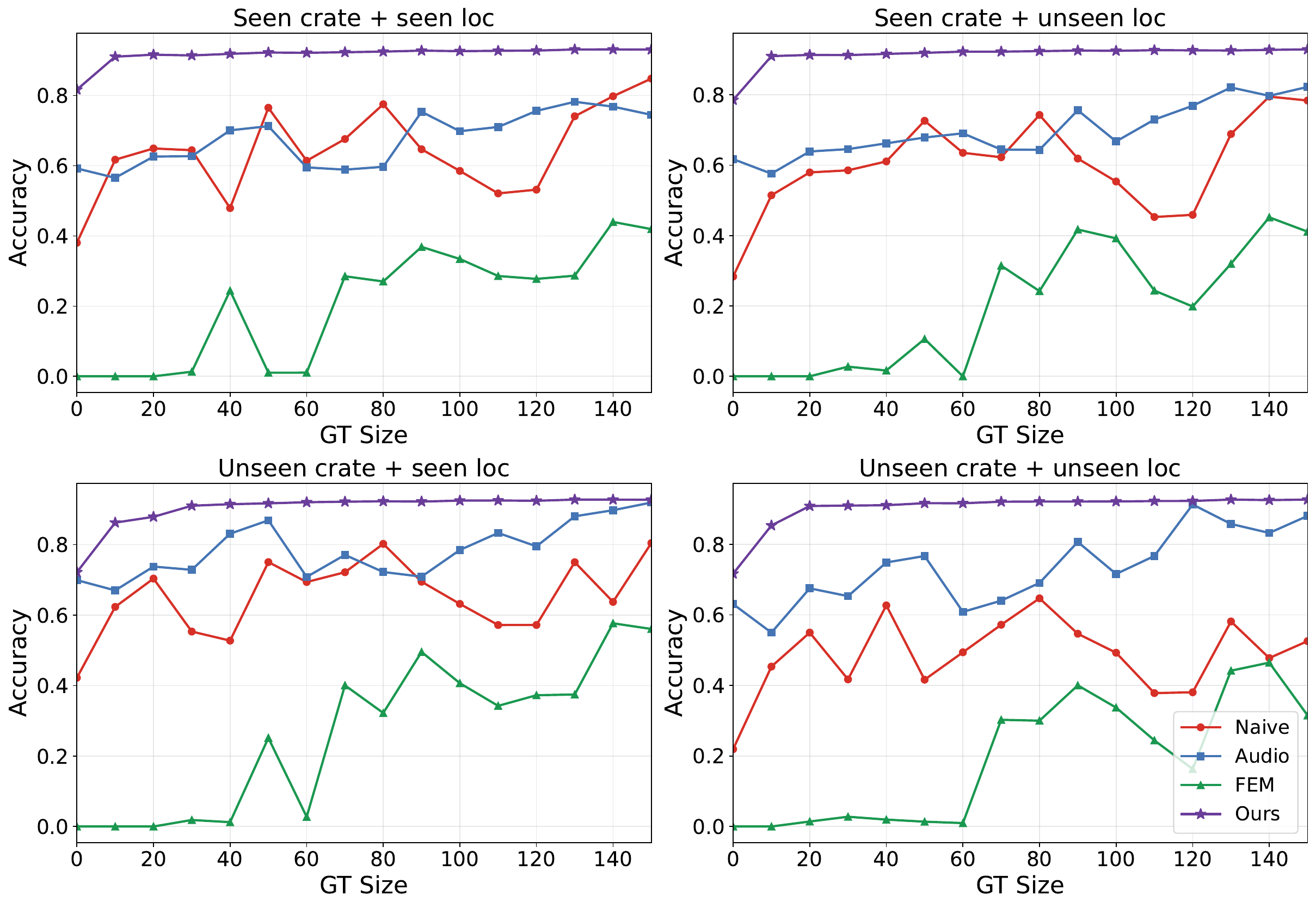}
    \caption{performance shift after finetune}
    \label{fig:performance_after_finetune}
\end{figure}

For rapid real-world deployment to be truly practical, the synthesized representations must not only exhibit competent zero-shot generalization out of the box, but also serve as an effective inductive prior that rapidly adapts as newly collected calibration samples accumulate over time. In this experiment, we investigate the downstream classifier adaptation trajectory using the pre-trained weights established in Section~\ref{subsubsec:downstream_utility}, tracking accuracy gains as real world supervision gradually expands. In detail, from the 40-hour test split previously reserved in Section~\ref{subsubsec:downstream_utility}, we set aside 10~hours for continuous evaluation, while incrementally drawing adaptation samples from the remaining 30~hours. To preclude optimization collapse stemming from transient label imbalances during data accumulation, we define a balanced ``chunk'' containing an equal instance of all four target behavioral classes. This chunk serves as the discrete step unit for the \textit{GT Size} metric in Figure~\ref{fig:performance_after_finetune} (e.g., a GT Size of 20 corresponds to 80 discrete training instances).

As illustrated in Figure~\ref{fig:performance_after_finetune}, our proposed framework demonstrates rapid and monotonic convergence, attaining the empirical performance ceiling ($\sim$92\% across all environments) with as few as 30 to 50 ground-truth chunks. In comparison, the terminal classification accuracies for the Naive, Audio, and FEM baselines level off at ranging 71.7\%-81.9\% across environments, 74.5\%-91.9\% across environments, and 46.2\%-56.1\% across environments, respectively. Baselines such as Naive and Audio exhibit markedly sluggish adaptation curves, with Audio hovering around or below 50\% overall accuracy for the first 110 adaptation steps. Notably, the FEM baseline yields virtually zero discernable performance gains across the target environments until reaching a GT Size of approximately 60. This protracted latency stems from the acute distributional discrepancy between FEM simulations and real physical boundary conditions, necessitating substantial in-domain supervision merely to override and unlearn the flawed synthetic prior before effective adaptation can begin.

\subsection{System Robustness}
\label{subsec:system_robustness}
While \system{} demonstrates strong end-to-end performance under canonical estimation scenarios, real-world deployment necessitates rigorous verification of system robustness under non-ideal, degraded operating regimes. Guided by elastodynamic principles, \system{} inherently couples excitation sources with structural geometries and medium properties, allowing it to leverage physical priors even when perceptual inputs are compromised. 

To demonstrate that this physics-grounded formulation prevents performance collapse, we subject \system{} to stress tests across two practical edge-case dimensions:
\begin{itemize}
    \item \textbf{Visual Occlusion (\ref{subsubsec:occlusion_scarcity}):} We evaluate scenarios where visual inputs are obscured by physical obstacles (\textit{Visual Occlusion}) or lack motion dynamics entirely due to unsynchronized capture or sensor dropouts (\textit{Temporal Scarcity}), proving that static structural cues and implicit material priors sustain reconstruction.
    \item \textbf{Unseen Dynamic Generalization (\ref{subsubsec:unseen_dynamics}):} We examine whether \system{} can reliably generalize to novel, unobserved behavioral excitations and across unseen structural/sensor transfer regimes, demonstrating that the framework learns governing propagation dynamics rather than memorizing training distributions.
\end{itemize}

\subsubsection{Visual Occlusion}
\label{subsubsec:occlusion_scarcity}
In practical environments, relying on time-series visual feeds presents inherent sensing vulnerabilities: 2D line-of-sight views are prone to physical occlusions, and multi-modal synchronization issues or data collection constraints often cause temporal visual scarcity where continuous motion video is unavailable. To assess whether \system{} can withstand such perceptual degradations, we identified instances as occluded when the ground-truth activity of the sow and piglets was labeled as non-idle (i.e., actively performing an action), yet no apparent motion was observed in the visual feed via Dense Inverse Search (DIS) optical flow. We then evaluated classification accuracy on this subset.

\begin{figure}[htbp]
\centering
    \includegraphics[width=0.99\linewidth]{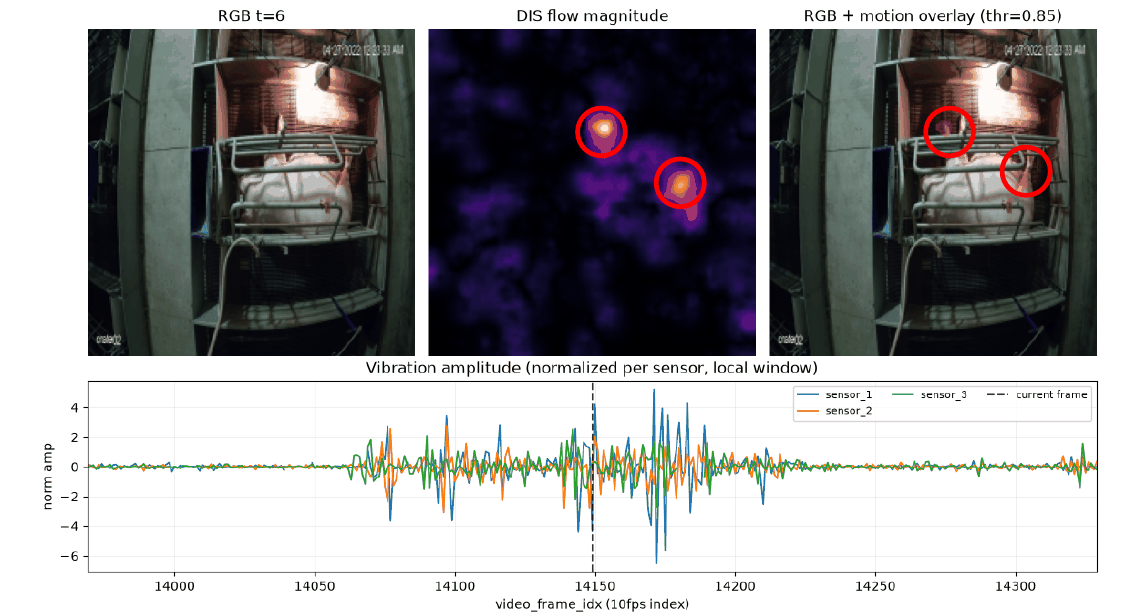}
    \caption{Vibration waveform and optical flow magnitude to capture pig's movement}
    \label{fig:excitation_visualziation}
\end{figure}

As illustrated in Figure~\ref{fig:excitation_visualziation}, the movement of pigs is captured well by DIS optical flow under standard conditions, closely aligning with the vibration signal amplitudes. In contrast, in the majority of occluded scenarios, the motion, primarily the sow's foot movement, is obscured behind the metal pipes of the farrowing crate. To reliably detect motion absence, we established a conservative threshold by selecting the maximum DIS optical flow value among all timestamps labeled with non-idle activities that indicated perceived motion. Under this criterion, occlusions accounted for a total of only 201.88 seconds across the entire dataset. Furthermore, these occluded intervals occurred transiently, spanning less than one second ($< 1\,\text{s}$) within any individual labeled instance. To address the scarcity of occlusion samples and construct a sufficient training dataset, we augmented the instances by applying sliding temporal windows around each occlusion event, shifting temporal offsets by 1-second and 5-second margins both backward and forward.

\begin{figure}[htbp]
\centering
    \includegraphics[width=0.90\linewidth]{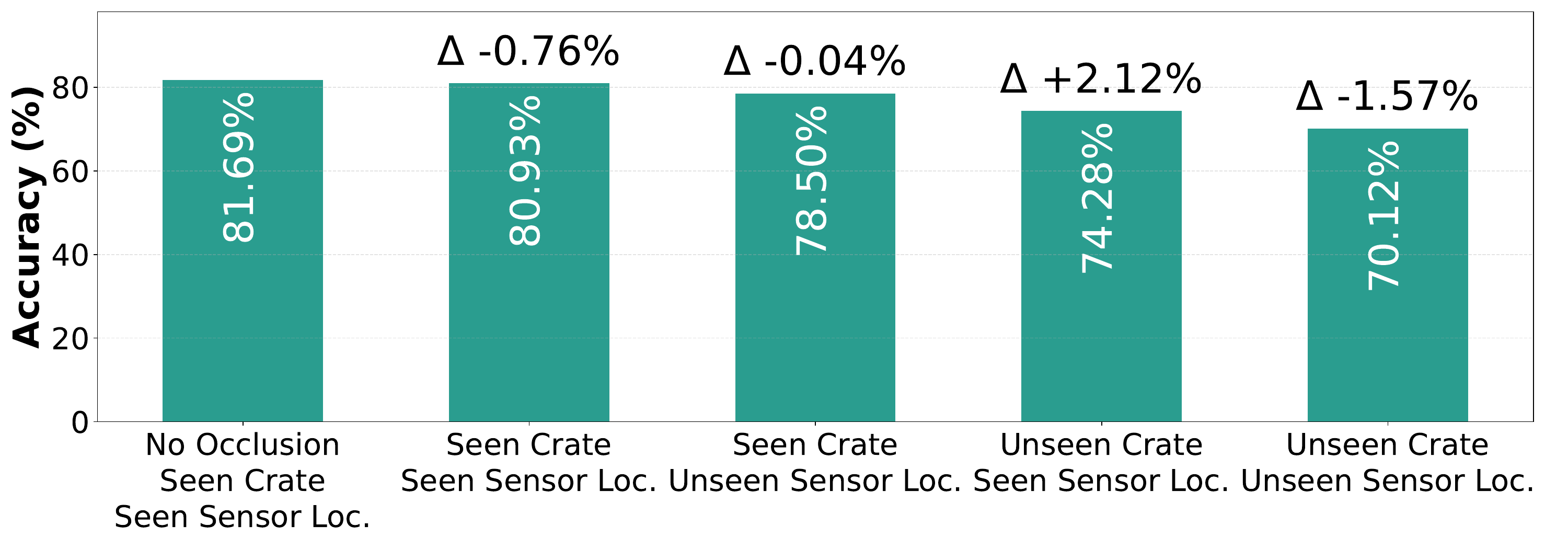}
    \caption{Occlusion by environment.}
    \label{fig:occlusion_by_env_plot}
\end{figure}

As shown in Figure~\ref{fig:occlusion_by_env_plot}, performance was barely degraded by the presence of visual occlusions compared to the unoccluded baseline (Seen Crate + Seen Sensor Loc., 81.69\%). Under the Seen Crate + Seen Sensor Loc. setup, the accuracy reached 80.93\%, showing a minor decrease of only 0.76\%. Similarly, for Seen Crate + Unseen Sensor Loc., accuracy remained resilient at 78.50\% ($\Delta -0.04\%$). In cross-crate environments, the accuracy achieved 74.28\% ($\Delta +2.12\%$) for Unseen Crate + Seen Sensor Loc., and 70.12\% ($\Delta -1.57\%$) for Unseen Crate + Unseen Sensor Loc. These results demonstrate that \system{} maintains robust classification capabilities across diverse environments under typical physical occlusions. To investigate more degradation scenarios, such as complete video freezing or absence of visual streams, we extend our evaluations in Section~\ref{subsec:no_vis_data}.

\subsubsection{Unseen Dynamic Generalization}
\label{subsubsec:unseen_dynamics}
To rigorously substantiate that \system{} internalizes authentic domain discrepancies across disparate deployment environments (\textit{seen/unseen crate} $\times$ \textit{seen/unseen sensor location}) while retaining expressive capacity for downstream inference, we conduct a joint geometric and spectral analysis using Linear Discriminant Analysis (LDA) and Feature Covariance Eigenspectrum. 
A well-generalized physical augmentation framework must satisfy two non-trivial criteria simultaneously: (i) it must discern physical structural and spatial domain shifts rather than collapsing into generic, location-agnostic artifacts, and (ii) this domain discrimination must not arise from superficial shortcuts or over-simplifications that sacrifice the nuanced physical dynamics required for downstream tasks.

\begin{figure}[htbp]
\centering
    \includegraphics[width=\linewidth]{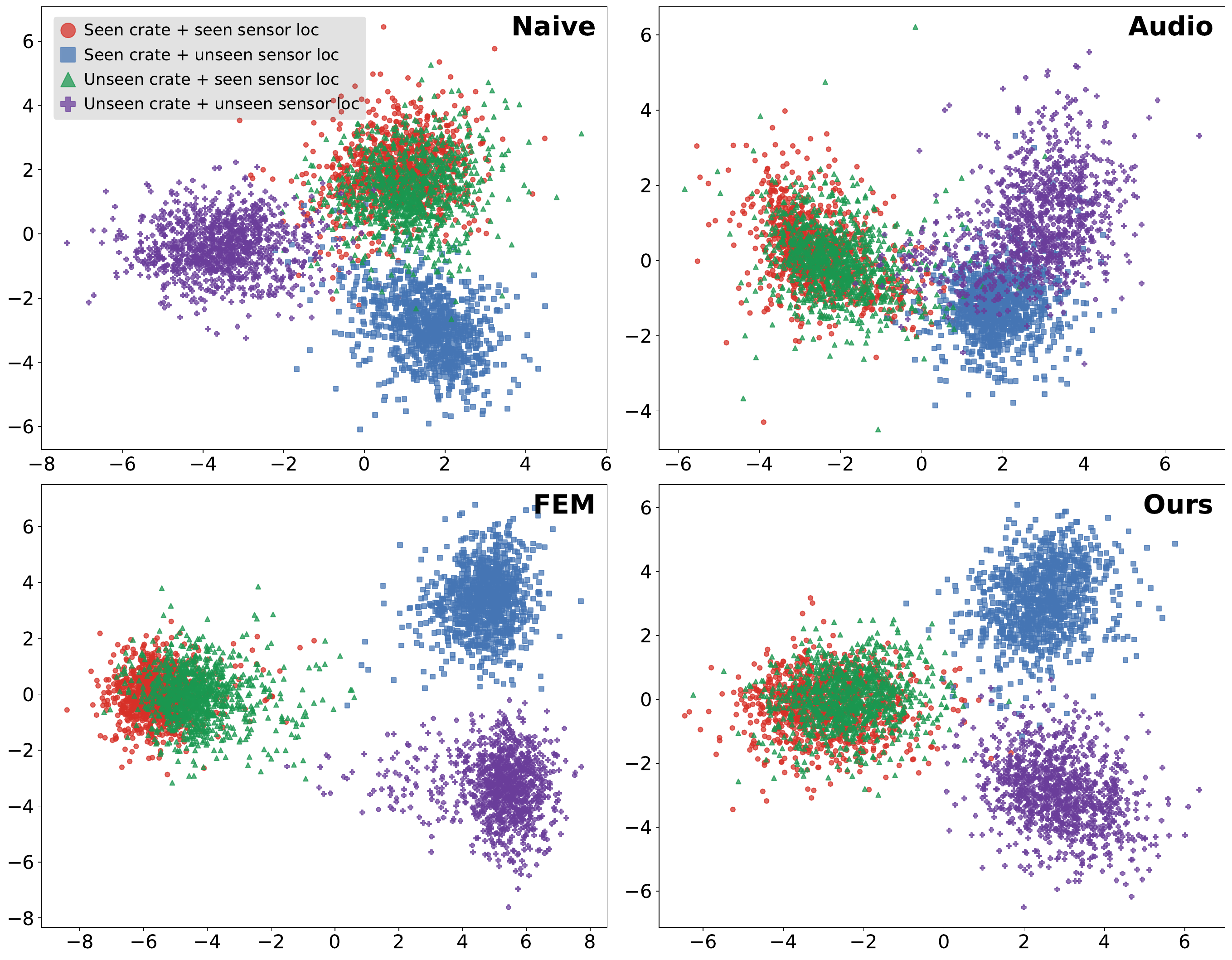}
    \caption{LDA across methods on 4 different environments}
    \label{fig:lda}
\end{figure}

To evaluate whether the synthesized representations preserve clear physical boundaries across deployment regimes, we employ Linear Discriminant Analysis (LDA). While unsupervised projections such as t-SNE or PCA capture arbitrary manifold variances or local neighborhood distortions without explicit label awareness, LDA computes supervised linear projections that explicitly maximize between-class variance ($S_B$) relative to within-class variance ($S_W$), directly quantifying whether domain-specific boundaries are linearly separable.
As depicted in Figure~\ref{fig:lda}, the four deployment environments exhibit starkly contrasting distributions across paradigms:
\begin{itemize}
    \item \textbf{Naive Data-Driven and Audio Encoder}: The Naive baseline clusters reasonably well within familiar settings, but completely collapses upon encountering the \textit{unseen crate}, producing overlapping, degenerate clusters that fail to establish distinct domain boundaries. Similarly, the Audio Encoder exhibits distorted, overlapping elliptical geometries across environments, demonstrating that airborne acoustic priors cannot reliably resolve solid-medium boundary variations.
    \item \textbf{Physics-Based FEM Simulation}: The FEM baseline yields sharp, isolated clusters with minimal intra-domain spread. However, this seemingly perfect linear separability is an artifact of over-idealized boundary conditions: because FEM relies on deterministic, noiseless equations governed by a minimal set of modal frequencies, it artificially forces representations into rigid, separated points.
    \item \textbf{\system{} (Ours)}: In contrast, \system{} achieves compact within-domain clustering while maintaining well-defined separation across all four transfer regimes. This confirms that our framework genuinely learns how elastic wave propagation operators transform under altered boundary conditions, without collapsing into mode mixtures.
\end{itemize}

\begin{figure}[htbp]
\centering
    \includegraphics[width=0.95\linewidth]{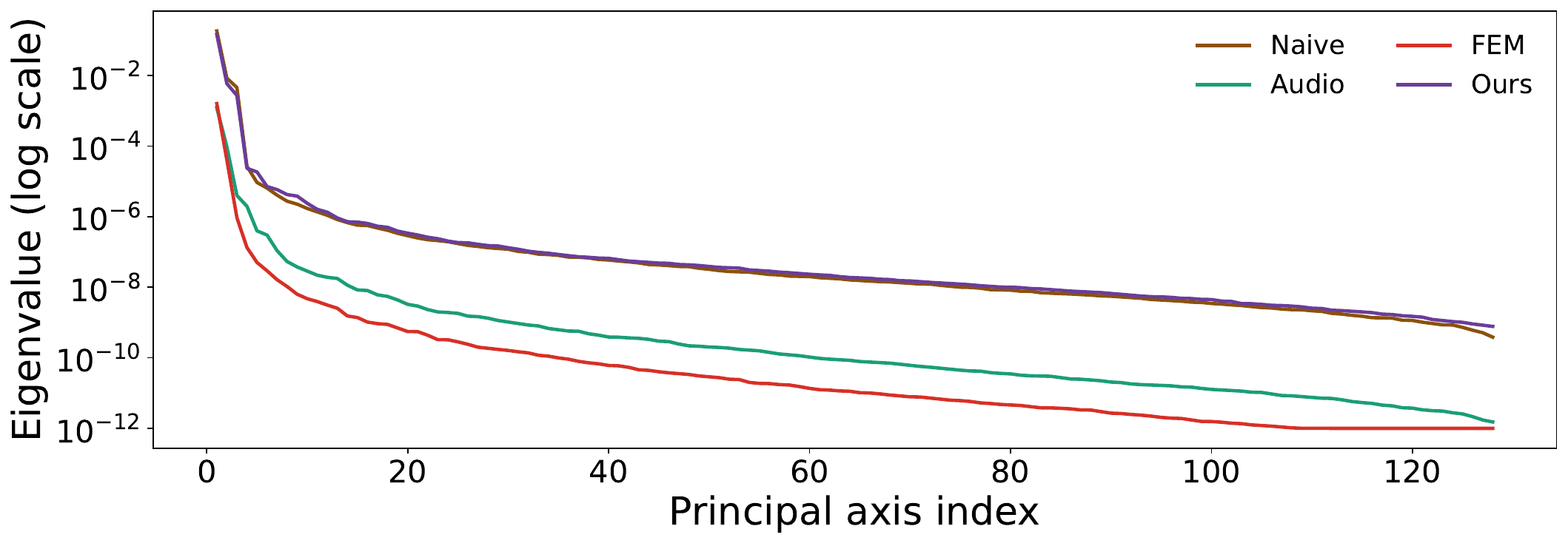}
    \caption{Feature covariance eigenspectrum across principal axis indices (log scale).}
    \label{fig:eigenspectrum_log}
\end{figure}

However, evaluating linear separability in isolation risks misinterpreting rigid, collapsed representations, such as those of FEM, as superior. To decouple true expressive capacity from synthetic dimensional collapse, we evaluate the eigenspectrum of the feature covariance matrices in Figure~\ref{fig:eigenspectrum_log}. 
The eigenspectrum scree plot directly reveals the effective dimensionality and residual energy across orthogonal feature axes:
\begin{itemize}
    \item The FEM baseline experiences catastrophic dimensional collapse, where the cumulative variance reaches 99\% within merely two principal axes ($k_{99}=2$, effective rank $\approx 1.13$), with subsequent eigenvalues dropping precipitously below $10^{-8}$. This mathematical collapse explains why FEM achieves trivial separation in LDA yet fails catastrophically (0.0\% accuracy) on downstream tasks: it strips away the fine-grained physical dynamics, contact transients, and damping variations essential for real-world inference.
    \item Conversely, although the Naive baseline exhibits a heavy eigenvalue tail comparable to \system{}, its poor LDA separability demonstrates that this high spectral energy stems from unstructured, isotropic noise rather than organized physical variance.
    \item Meanwhile, the Audio baseline suffers from partial spectral decay due to modality misalignment, yet its pre-trained wave dynamics enable it to retain superior downstream utility over the unstructured Naive model.
\end{itemize}

\system{} is the only paradigm that reconciles this fundamental tension. It maintains high spectral capacity ($k_{99}=3$, effective rank $\approx 1.29$) across higher-order principal axes, sustaining several orders of magnitude higher energy than FEM and Audio throughout the spectrum, while concurrently enforcing structured domain separability in LDA space. This joint analysis proves that \system{} neither memorizes generic shortcuts nor sacrifices rich latent physical dynamics, establishing an optimal balance that yields superior transferability and downstream generalization.

As illustrated in Figure~\ref{fig:lda}, a shared geometric tendency across all evaluated paradigms is that representations for \textit{Unseen Crate + Seen Sensor Loc.} and \textit{Seen Crate + Seen Sensor Loc.} are projected closely together. This empirical alignment indicates that physical relocation of the sensor induces a substantially more distributional distortion in elastodynamic wavefields than global structural variations of the crate itself. Corroborating this physical property, the two configurations sharing unseen sensor placements, namely, \textit{Seen Crate + Unseen Sensor Loc.} (blue) and \textit{Unseen Crate + Unseen Sensor Loc.} (purple), likewise exhibit closer proximity to each other while remaining distinct from the seen-sensor regimes.

In terms of environment-level discriminability, noticeable disparities emerge across baselines. In particular, the \textbf{Audio Encoder} baseline exhibits the weakest domain separation, failing to isolate the four deployment environments into distinct geometric manifolds. Conversely, the \textbf{FEM} baseline achieves exceptionally sharp boundaries between the four settings. Meanwhile, the \textbf{Naive Data-Driven} baseline displays slightly more clustered and compressed representations than \system{}. This implies that, with the exception of the Audio model, all methods internalize input conditions well enough to differentiate disparate physical settings to some extent. However, the inter-environment margin of the Naive baseline remains visibly weak and diffuse, preventing it from establishing clear-cut domain decision boundaries.

While the LDA projection in Figure~\ref{fig:lda} evaluates whether a model recognizes domain-specific physical shifts, the feature covariance eigenspectrum in Figure~\ref{fig:eigenspectrum_log} examines whether the latent representations retain sufficient expressive capacity to facilitate downstream tasks rather than merely preserving trivial classification shortcuts. In an eigenspectrum scree plot, sustaining high eigenvalue magnitudes across higher-order principal axes (i.e., as the axis index increases toward the right) indicates a richer, higher-dimensional latent representation that resists dimensional collapse. 

Because the FEM baseline demonstrated near-perfect environmental separation in Figure~\ref{fig:lda} yet suffered a performance collapse in our downstream evaluation (Section~\ref{subsubsec:downstream_utility}), validating its spectral dimensionality is critical. As revealed in Figure~\ref{fig:eigenspectrum_log}, the FEM features yield the lowest eigenvalues across the entire principal axis spectrum among all four evaluated paradigms. This drastic spectral decay confirms that the synthetic data generated by FEM discards essential micro-scale physical dynamics and contact transients required for practical downstream inference.

In the case of the \textbf{Audio Encoder}, the model exhibits poor environmental separability while retaining an eigenspectrum only marginally richer than that of FEM. Nevertheless, as observed in Section~\ref{subsubsec:downstream_utility}, it achieves the second-highest downstream classification accuracy behind \system{}. This indicates that the pre-trained audio backbone effectively captures core oscillatory and temporal-frequency priors that are indispensable for downstream classification, remaining resilient across structural shifts. However, because its representation is grounded strictly in airborne acoustic corpora rather than solid-state elastodynamics, it encounters an intrinsic ceiling when adapting to complex structural boundary variations.

Finally, while the \textbf{Naive Data-Driven} baseline tracks the eigenvalue spectrum of \system{} almost identically across all principal indices, suggesting an ostensibly heavy feature capacity, its inferior separability in Figure~\ref{fig:lda} reveals that this high spectral energy stems primarily from unconstrained, task-irrelevant noise rather than an authentic understanding of elastodynamic propagation.

\section{Ablation}
\label{sec:Ablation}

Throughout Section~\ref{sec:Evaluation}, we demonstrated the end-to-end performance of our system and its robustness to unseen crates and sensor locations. We also analyzed the underlying factors contributing to the results across different methods. 

In this section, we conduct ablation studies by imposing constraints on specific modules of our system to quantify their individual contributions.

\subsection{Impact of Dynamic Visual Information}
\label{subsec:no_vis_data}

A key distinction between our system and conventional synthetic data generation approaches is the extraction of physical dynamics from consecutive video frames, which provides supplementary contextual information to the model. Therefore, validating the practical efficacy of such visual guidance is critical. 

To evaluate this, we modify the original \system{} pipeline by replacing the video input with a static sequence: the initial frame of each video is duplicated across the entire duration. This setup allows the model to extract static structural information from the scene while preventing it from observing real-time interactions of objects or animals. Consequently, the network cannot compute the delta energy values via ResNet, nor can it capture structural variations caused by pig locomotion. Under this constraint, the model must rely solely on the input vibration signals to generate the corresponding STFT representations.

\begin{figure}[htbp]
\centering
    \includegraphics[width=0.9\linewidth]{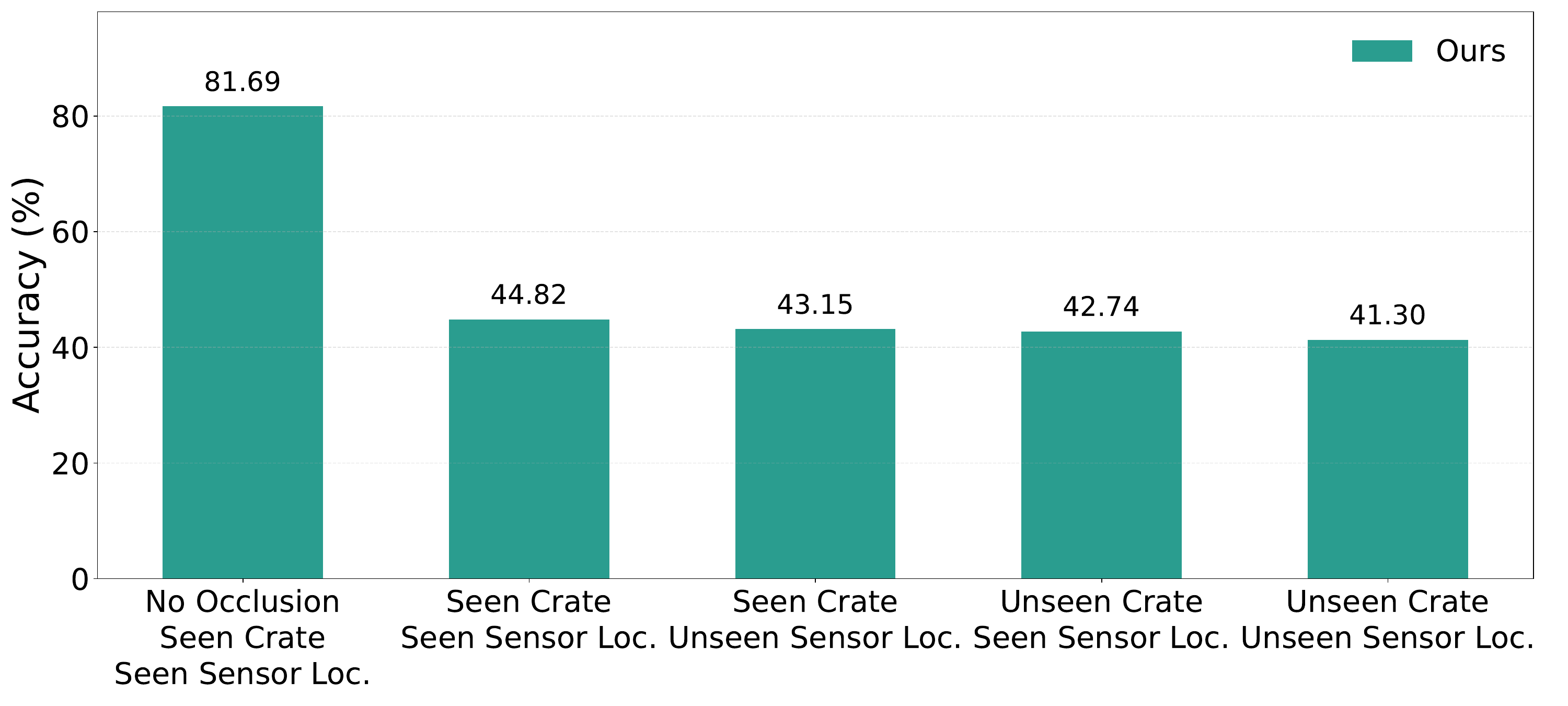}
    \caption{Performance comparison under static image inputs across different environments.}
    \label{fig:static_image_by_env_plot}
\end{figure}

As illustrated in Figure~\ref{fig:static_image_by_env_plot}, the system's performance (represented by the green bar) deteriorates substantially compared to the unoccluded baseline shown on the far left. This performance degradation indicates that capturing dynamic structural changes, particularly the delta energy derived from video sequences, is crucial for synthesizing STFT features that preserve sufficient representational capacity to ensure high performance on downstream tasks.

\subsection{Performance by Module}

\system{} projects the visual inputs into 3D space via the Propagation Possibility Map to identify candidate regions with high propagation likelihood, while leveraging the Visual Feature Map to capture source kinetics (via the temporal feature-difference descriptor $E_{\Delta}(t)$) and implicit material dynamics (via GAP and GMP). To validate the complementary nature and indispensability of these two components, we conducted an ablation study by training models that utilize each feature independently.

For the \textit{Visual Feature Map Only} variant, we utilized the initial visual feature representations learned during the standard training pipeline of \system{}. For the \textit{Propagation Map Only} variant, we isolated the propagation branch, exploiting the structurally decoupled architecture described in Section~\ref{sec:s2s_training}. Both baseline models were trained using the same dataset as Stage~1 of the proposed method (\system{}). The training configuration remained identical, with models trained for up to 100 epochs using an early stopping patience of 3 and retaining the checkpoint from the best epoch.

\begin{table}[h]
\centering
\caption{Downstream task performance evaluation across different crate and sensor location settings (Loc. = Sensor Location).}
\label{tab:ablation_downstream_results}
\resizebox{\linewidth}{!}{%
\begin{tabular}{lcccccccc}
\toprule
\multirow{2}{*}{\textbf{Method}} & \multicolumn{2}{c}{\textbf{Seen Crate + Seen Loc.}} & \multicolumn{2}{c}{\textbf{Seen Crate + Unseen Loc.}} & \multicolumn{2}{c}{\textbf{Unseen Crate + Seen Loc.}} & \multicolumn{2}{c}{\textbf{Unseen Crate + Unseen Loc.}} \\
\cmidrule(lr){2-3} \cmidrule(lr){4-5} \cmidrule(lr){6-7} \cmidrule(lr){8-9}
& Acc. (\%) $\uparrow$ & Micro-F1 $\uparrow$ & Acc. (\%) $\uparrow$ & Micro-F1 $\uparrow$ & Acc. (\%) $\uparrow$ & Micro-F1 $\uparrow$ & Acc. (\%) $\uparrow$ & Micro-F1 $\uparrow$ \\
\midrule
Proposed method (\system{}) & 81.69\% & 0.8180 & 78.54\% & 0.7437 & 72.16\% & 0.6852 & 71.69\% & 0.6149 \\
Visual Feature Map Only     & 64.80\% & 0.6321 & 62.54\% & 0.7107 & 58.71\% & 0.5752 & 59.20\% & 0.5471 \\
Propagation Map Only        & 29.58\% & 0.2717 & 29.56\% & 0.1492 & 32.21\% & 0.2763 & 28.57\% & 0.2682 \\
\bottomrule
\end{tabular}%
}
\end{table}

As presented in Table~\ref{tab:ablation_downstream_results}, the \textit{Visual Feature Map Only} configuration outperforms the \textit{Propagation Map Only} model by a substantial margin. This suggests that even coarse spatial and dynamic cues provide significant supervisory signals for vibration reconstruction. Conversely, utilizing the propagation map alone proves insufficient for accurate vibration analysis. This performance drop can be attributed to the inherent probabilistic nature and uncertainty embedded in the propagation map, which poses a substantial challenge for the network to infer deterministic propagation dynamics independently. Crucially, when both modules are unified within \system{}, their synergistic interaction yields substantial performance improvements, validating the efficacy of our joint design.

\section{Conclusion}
\label{sec:Conclusion}

In this paper, we tackled the fundamental challenge of spatial sensitivity in structure-borne vibration sensing, where physical reallocations and structural variations distort elastic wavefields and invalidate learned representations.

We introduced \system{}, an elastodynamically grounded spatial augmentation framework that leverages monocular visual context as soft spatial priors. By probabilistically formulating wave-guiding pathways via a Vision-Derived Propagation Possibility Map and extracting dynamic kinetic signatures through a Visual Feature Map, \system{} reliably synthesizes realistic vibration profiles across unmeasured coordinates from a single reference sensor. Extensive empirical evaluations across 1,237 sensor-hours from real-world deployments confirm that \system{} consistently outperforms heuristic, generative, and physics-based FEM baselines in both signal reconstruction and downstream task transfer. Beyond signal fidelity, our framework substantially mitigates deployment friction: it eliminates the heavy human burden of collecting exhaustive ground-truth profiles from scratch whenever a sensor is relocated, enabling downstream classifiers to rapidly calibrate with minimal human intervention.

Looking ahead, the physics-consistent surrogate paradigm demonstrated in \system{} opens up effective avenues for human centered ubiquitous computing. By freeing structure-borne sensing from rigid physical geometries without demanding exhaustive engineering calibration, this approach directly generalizes to indoor human sensing domains—such as footstep-induced indoor localization, ambient fall detection, and smart workplace activity recognition. By bridging the gap between sparse sensing feasibility and full spatial observability, \system{} provides a practical stepping stone toward scalable, small-overhead spatial intelligence in dynamically inhabited physical spaces.   


\bibliographystyle{ACM-Reference-Format}
\bibliography{references, eis-lab}

\clearpage
\appendix
\section{Appendix}
\label{sec:appendix}

\subsection{Model Parameter Counts across Architectures and Scales}

\begin{table}[t]
\centering
\caption{Model Parameter Counts across Architectures and Scales.}
\label{tab:model_parameters}
\begin{tabular}{llrr}
\toprule
\textbf{Architecture} & \textbf{Size} & \textbf{Vibration Params} & \textbf{Vision Params} \\
\midrule
\multirow{3}{*}{CNN} 
& small  & 141,445   & 99,589    \\
& medium & 561,413   & 391,685   \\
& large  & 2,236,933 & 1,553,413 \\
\midrule
\multirow{3}{*}{RNN} 
& small  & 125,445   & 183,237   \\
& medium & 793,093   & 1,019,269 \\
& large  & 4,341,765 & 5,236,485 \\
\midrule
\multirow{3}{*}{TCN} 
& small  & 50,245    & 107,909   \\
& medium & 297,349   & 523,269   \\
& large  & 1,578,245 & 2,472,453 \\
\midrule
\multirow{3}{*}{Transformer} 
& small  & 102,085   & 158,597   \\
& medium & 599,045   & 822,661   \\
& large  & 3,167,493 & 4,057,093 \\
\bottomrule
\end{tabular}
\end{table}

\subsection{FEM Setting}
\label{subsec:fem_setting}

To model elastic wave propagation across the structure and reconstruct unknown vibration sources, we discretize the 3D domain using the Finite Element Method (FEM) via SfePy. When precise alloy grades or spatial variations of composite assemblies remain unidentified, approximating the structure as homogeneous structural steel provides a physically sound and consistent engineering baseline. Wave propagation velocities depend primarily on the ratio $\sqrt{E/\rho}$; hence, structural geometry predominantly dictates modal characteristics.

The constitutive relation is governed by the isotropic elasticity tensor $\mathbf{D} \in \mathbb{R}^{6 \times 6}$ defined via Lam{\'e} parameters:
\begin{equation}
    \lambda = \frac{E \nu}{(1 + \nu)(1 - 2\nu)}, \quad \mu = \frac{E}{2(1 + \nu)},
\end{equation}
\begin{equation}
    \mathbf{D} = \begin{bmatrix}
        \lambda + 2\mu & \lambda & \lambda & 0 & 0 & 0 \\
        \lambda & \lambda + 2\mu & \lambda & 0 & 0 & 0 \\
        \lambda & \lambda & \lambda + 2\mu & 0 & 0 & 0 \\
        0 & 0 & 0 & \mu & 0 & 0 \\
        0 & 0 & 0 & 0 & \mu & 0 \\
        0 & 0 & 0 & 0 & 0 & \mu
    \end{bmatrix}.
\end{equation}

Table~\ref{tab:fem_settings} summarizes the material constants, discretization requirements, and numerical solver configurations.

\begin{table}[htbp]
\centering
\caption{FEM Simulation Configurations and Structural Steel Parameters}
\label{tab:fem_settings}
\begin{tabular}{lll}
\toprule
\textbf{Configuration Category} & \textbf{Design Specification} & \textbf{Value / Choice} \\
\midrule
Material Density ($\rho$) & Structural Steel Baseline & $7850\text{ kg/m}^3$ \\
Young's Modulus ($E$) & Baseline Elasticity & $200.0\text{ GPa}$ \\
Poisson's Ratio ($\nu$) & Transverse Contraction & $0.30$ \\
Rayleigh Damping ($\alpha, \beta$) & Mass/Stiffness Damping & $0.05\text{ s}^{-1},\; 1.0\times 10^{-5}\text{ s}$ \\
Spatial Element Type & 3D Tetrahedron & 10-node quadratic tetrahedral \\
Polynomial Order ($p$) & Quadratic Basis & $p = 2$ (Mitigate shear locking) \\
Elements per Wavelength & Dispersion Control & $N_{\lambda} \ge 8$ at $f_{\max}$ \\
Time-Marching Scheme & Implicit Newmark-$\beta$ & $\beta = 0.25, \; \gamma = 0.50$ \\
Sampling Step ($\Delta t$) & Temporal Discretization & $\le 1 / (20 f_{\max})$ \\
Boundary Condition ($\Gamma_{\text{base}}$) & Fixed Support (Dirichlet) & $\mathbf{u}|_{\Gamma_{\text{base}}} = \mathbf{0}$ \\
\bottomrule
\end{tabular}
\end{table}

\subsubsection{Discretization and Numerical Convergence}
To capture thin-walled pipe bending without artificial stiffening, cross-sections are discretized with at least $2$--$3$ tetrahedral element layers. Quadratic ($p=2$) basis functions are strictly enforced to eliminate shear locking. Numerical wave dispersion is controlled by restricting element size to $h \le \lambda_{\min} / 8 = c_s / (8 f_{\max})$, where $c_s = \sqrt{\mu / \rho}$. For transient integration, an unconditionally stable implicit Newmark-$\beta$ method is employed with time step $\Delta t \le 1 / (20 f_{\max})$. Mesh convergence is verified by confirming a frequency drift $|\omega_r^{(h)} - \omega_r^{(h/2)}| / \omega_r^{(h/2)} \le 1.0\%$ across the primary elastic modes.

\subsubsection{Modal Extraction and Inverse Source Identification}
Instead of solving forward transient systems iteratively for each candidate point, system responses are evaluated using modal superposition. Algorithm~\ref{alg:sfepy_modal} assembles the structural matrices and extracts mass-normalized modes. Algorithm~\ref{alg:source_inversion} computes the regularized cross-transfer ratios across candidate contact locations and optimizes their energy weights via Non-Negative Least Squares (NNLS) to predict the vibration waveform at uninstrumented locations.

\begin{algorithm}[htbp]
\caption{FEM Matrix Assembly and Modal Analysis via SfePy}
\label{alg:sfepy_modal}
\begin{algorithmic}[1]
\Require Volumetric mesh $\Omega$, Elastic tensor $\mathbf{D}(E, \nu)$, Density $\rho$, Foundation support threshold $z_{\text{thresh}}$, Truncation mode count $K_{\text{modes}}$.
\Ensure Eigenfrequencies $\{\omega_r\}_{r=1}^{K_{\text{modes}}}$, Mass-orthonormal mode matrix $\boldsymbol{\Phi} \in \mathbb{R}^{N_{\text{dof}} \times K_{\text{modes}}}$.
\State Identify base boundary: $\Gamma_{\text{base}} \gets \{ \mathbf{x} \in \partial\Omega \mid z(\mathbf{x}) \le z_{\text{thresh}} \}$.
\State Construct displacement vector field $\mathbf{u}(\mathbf{x}) \in [\mathcal{H}^1(\Omega)]^3$ with quadratic basis ($p=2$).
\State Assemble global stiffness $\mathbf{K}$ and mass $\mathbf{M}$ matrices:
    \Statex \hspace{\algorithmicindent} $\mathbf{K}_{ij} \gets \int_\Omega \boldsymbol{\varepsilon}(\mathbf{v}_i) : \mathbf{D} : \boldsymbol{\varepsilon}(\mathbf{u}_j) \, d\Omega$
    \Statex \hspace{\algorithmicindent} $\mathbf{M}_{ij} \gets \int_\Omega \rho \, \mathbf{v}_i \cdot \mathbf{u}_j \, d\Omega$
\State Enforce Dirichlet constraints: $\mathbf{u}|_{\Gamma_{\text{base}}} = \mathbf{0}$.
\State Solve generalized symmetric eigenvalue problem using shift-invert Lanczos at $\sigma = 0$:
    \Statex \hspace{\algorithmicindent} $\mathbf{K} \boldsymbol{\phi}_r = \omega_r^2 \mathbf{M} \boldsymbol{\phi}_r, \quad \text{s.t.} \quad \boldsymbol{\phi}_r^T \mathbf{M} \boldsymbol{\phi}_s = \delta_{rs}$
\State Sort pairs such that $0 < \omega_1 \le \omega_2 \le \dots \le \omega_{K_{\text{modes}}}$.
\State \Return $\{\omega_r\}_{r=1}^{K_{\text{modes}}}$ and $\boldsymbol{\Phi} = [\boldsymbol{\phi}_1, \dots, \boldsymbol{\phi}_{K_{\text{modes}}}]$.
\end{algorithmic}
\end{algorithm}

\begin{algorithm}[htbp]
\caption{Contact Source Inversion and Target Response Synthesis}
\label{alg:source_inversion}
\begin{algorithmic}[1]
\Require Mode shapes $\boldsymbol{\Phi}$, Frequencies $\{\omega_r\}$, Candidate source nodes $\mathcal{S}_{\text{cand}}$, Reference sensor node $n_{\text{in}}$, Target sensor node $n_{\text{out}}$, Modal damping $\zeta$, Measured signal $x_{\text{in}}(t)$, Empirical transfer function $H_{\text{emp}}(\omega)$.
\Ensure Estimated source weights $\mathbf{w}^*$, Predicted response waveform $\hat{x}_{\text{out}}(t)$.
\For{each discrete frequency $\omega_k \in \boldsymbol{\Omega}_{\text{eval}}$}
    \For{each candidate source node $s \in \mathcal{S}_{\text{cand}}$}
        \State Compute modal receptances to observation nodes:
            \Statex \hspace{\algorithmicindent}\hspace{\algorithmicindent} $G_{\text{in}}(s, \omega_k) \gets \sum_{r=1}^{K_{\text{modes}}} \frac{\phi_r(n_{\text{in}}) \phi_r(s)}{\omega_r^2 - \omega_k^2 + 2j\zeta\omega_r\omega_k}$
            \Statex \hspace{\algorithmicindent}\hspace{\algorithmicindent} $G_{\text{out}}(s, \omega_k) \gets \sum_{r=1}^{K_{\text{modes}}} \frac{\phi_r(n_{\text{out}}) \phi_r(s)}{\omega_r^2 - \omega_k^2 + 2j\zeta\omega_r\omega_k}$
        \State Formulate source-specific regularized transfer ratio:
            \Statex \hspace{\algorithmicindent}\hspace{\algorithmicindent} $T_s(\omega_k) \gets \frac{G_{\text{out}}(s, \omega_k) \cdot G_{\text{in}}^*(s, \omega_k)}{|G_{\text{in}}(s, \omega_k)|^2 + \epsilon_{\text{reg}}}$
    \EndFor
\EndFor
\State Stack real and imaginary transfer operators across frequencies:
    \Statex \hspace{\algorithmicindent} $\mathbf{A} \gets \begin{bmatrix} \text{Re}(\mathbf{T}) \\ \text{Im}(\mathbf{T}) \end{bmatrix} \in \mathbb{R}^{2N_f \times |\mathcal{S}_{\text{cand}}|}, \quad \mathbf{b} \gets \begin{bmatrix} \text{Re}(H_{\text{emp}}) \\ \text{Im}(H_{\text{emp}}) \end{bmatrix} \in \mathbb{R}^{2N_f}$
\State Optimize candidate energy weights via non-negative least squares:
    \Statex \hspace{\algorithmicindent} $\mathbf{w}^* \gets \arg\min_{\mathbf{w} \ge \mathbf{0}} \|\mathbf{A}\mathbf{w} - \mathbf{b}\|_2^2$
\State Normalize active weights: $\mathbf{w}^* \gets \mathbf{w}^* / \sum_m w_m^*$.
\State Synthesize composite transfer function: $T_{\text{syn}}(\omega) \gets \sum_{s} w_s^* T_s(\omega)$.
\State Predict target response spectrum: $\hat{X}_{\text{out}}(\omega) \gets \mathcal{F}\{x_{\text{in}}(t)\} \cdot T_{\text{syn}}(\omega)$.
\State Synthesize time waveform: $\hat{x}_{\text{out}}(t) \gets \mathcal{F}^{-1}\{\hat{X}_{\text{out}}(\omega)\}$.
\State \Return $\mathbf{w}^*$ and $\hat{x}_{\text{out}}(t)$.
\end{algorithmic}
\end{algorithm}

\end{document}